\documentclass{article}
\usepackage{amssymb}

\usepackage{amsmath}

\newtheorem{theorem}{Theorem}

\newtheorem{definition}[theorem]{Definition}
\newtheorem{example}[theorem]{Example}

\newtheorem{lemma}[theorem]{Lemma}

\begin{document}

\author{Petr Z\'avada \\
%EndAName
\\
\ \textit{Institute of Physics, Academy of Sciences of Czech Republic,}\\
\textit{Na Slovance 2, CZ-182 21 Prague 8}\\
e-mail: zavada@fzu.cz}
\date{May 9, 2001}
\title{Relativistic wave equations with fractional derivatives and
pseudo-differential operators}
\maketitle

\begin{abstract}
The class of the free relativistic covariant equations generated by the
fractional powers of the D'Alambertian operator $(\square ^{1/n})$ is studied.
Meanwhile the equations corresponding to $n=1$ and $2$ (Klein-Gordon and
Dirac equations) are local in their nature, the multicomponent equations for
arbitrary $n>2$ are non-local. It is shown, how the representation of
generalized algebra of Pauli and Dirac matrices looks like and how these
matrices are related to the algebra of $SU(n)$ group. The corresponding
representations of the Poincar\'e group and further symmetry transformations
on the obtained equations are discussed. The construction of the related
Green functions is suggested.
\end{abstract}

\numberwithin{equation}{section}

\section{Introduction}

The relativistic covariant wave equations represent an intersection of ideas
of the theory of relativity and quantum mechanics. The first and best known
relativistic equations, the Klein-Gordon and particularly Dirac equation,
belong to the essentials, which our present understanding of the microworld
is based on. In this sense it is quite natural, that the searching for and
the study of the further types of such equations represent a field of stable
interest. For a review see e.g. \cite{rev} and citations therein. In fact,
the attention has been paid first of all to the study of equations
corresponding to the higher spins $(s\geq 1)$ and to the attempts to solve
the problems, which have been revealed in the connection with these
equations, e.g. the acausality due to external fields introduced by the
minimal way.

In this paper we study the class of equations obtained by the
'factorization' of the D'Alambertian operator, i.e. by a generalization of
the procedure, by which the Dirac equation is obtained. As the result, from
each degree of extraction $n$ we get a multi-component equation, hereat the
case $n=2$ corresponds to the Dirac equation. However the equations for $n>2$
differ substantially from the cases $n=1,2$ since they contain fractional
derivatives (or pseudo-differential operators), so in the effect their
nature is non-local.

In the first part (Sec. 2), the generalized algebras of the Pauli and Dirac
matrices are considered and their properties are discussed, in particular
their relation to the algebra of the $SU(n)$ group. The second, main part
(Sec. 3) deals with the covariant wave equations generated by the roots of
the D'Alambertian operator, these roots are defined with the use of the
generalized Dirac matrices. In this section we show the explicit form of the
equations, their symmetries and the corresponding transformation laws. We
also define the scalar product and construct the corresponding Green
functions. The last section (Sec. 4) is devoted to the summary and
concluding remarks.

Let us remark, the application of the pseudo-differential operators in the
relativistic equations is nothing new. The very interesting aspects of the
scalar relativistic equations based on the square root of the Klein-Gordon
equation are pointed out e.g. in the papers \cite{suc}-\cite{smi}. Recently,
an interesting approach for the scalar relativistic equations based on the
pseudo-differential operators of the type $f(\square )$ has been proposed in
the paper \cite{bar}. One can mention also the papers \cite{szw}, \cite{ker}
in which the square and cubic roots of the Dirac equation were studied in
the context of supersymmetry. The cubic roots of the Klein-Gordon equation
were discussed in the recent papers \cite{ply}, \cite{ras}.

It should be observed, that our considerations concerning the generalized
Pauli and Dirac matrices (Sec. 2) have much common with the earlier studies
related to the generalized Clifford algebras (see e.g. \cite{ram}-\cite{tra}
and citation therein) and with the paper \cite{pat}, even if our starting
motivation is rather different.

\section{Generalized algebras of Pauli and Dirac matrices}

Anywhere in the next by the term \textit{matrix} we mean the square matrix $%
n\times n,$ if not stated otherwise. Considerations of this section are
based on the matrix pair introduced as follows.

\begin{definition}
\label{def1}For any $n\geq 2$ we define the matrices%
\begin{equation}
S=\left( 
\begin{array}{cccccc}
0 & \qquad & \qquad & \qquad & \qquad & 1 \\ 
1 &  &  &  &  & \qquad \\ 
\qquad & 1 &  &  &  &  \\ 
&  & \cdot &  &  &  \\ 
&  &  & \cdot &  &  \\ 
&  &  &  & 1 & 0%
\end{array}%
\right) ,  \label{ga1}
\end{equation}%
\begin{equation}
T=\left( 
\begin{array}{cccccc}
1 & \qquad & \qquad & \qquad & \qquad &  \\ 
\qquad & \alpha &  &  &  &  \\ 
&  & \alpha ^{2} &  &  &  \\ 
&  &  & \cdot &  &  \\ 
&  &  &  & \cdot &  \\ 
&  &  &  &  & \alpha ^{n-1}%
\end{array}%
\right)  \label{gb1}
\end{equation}%
where $\alpha =\exp (2\pi i/n)$ and in the remaining empty positions are
zeros.
\end{definition}

\begin{lemma}
Matrices $X=S,T$ satisfy the following relations 
\begin{equation}  \label{ga2}
\alpha ST=TS,
\end{equation}
\begin{equation}  \label{ga3}
X^n=I,
\end{equation}
\begin{equation}  \label{ga4}
XX^{\dagger }=X^{\dagger }X=I,
\end{equation}
\begin{equation}  \label{ga5}
\det X=(-1)^{n-1},
\end{equation}
\begin{equation}  \label{ga6}
\text{\textrm{Tr} }X^k=0,\qquad k=1,2...n-1,
\end{equation}
where $I$ denotes the unit matrix.
\end{lemma}

\noindent \emph{Proof:}

\noindent All the relations easily follow from the \textit{Definition \ref%
{def1}.}

\begin{definition}
\label{def2}Let $\ \mathcal{A}$ be some algebra on the field of complex
numbers, $(p,m)$ be a pair of natural numbers, $X_{1},X_{2},...,X_{m}\in 
\mathcal{A}$ and $a_{1},a_{2},...,a_{m}$ $\in C$. The $p-th$ power of the
linear combination can be expanded: 
\begin{equation*}
\left( \sum_{k=1}^{m}a_{k}X_{k}\right)
^{p}=\sum_{p_{j}}a_{1}^{p_{1}}a_{2}^{p_{2}}...a_{m}^{p_{m}}\left\{
X_{1}^{p_{1}},X_{2}^{p_{2}},...,X_{m}^{p_{m}}\right\} ;\quad
p_{1}+...+p_{m}=p,
\end{equation*}%
where the symbol $\{X_{1}^{p_{1}},X_{2}^{p_{2}},...,X_{m}^{p_{m}}\}$
represents the sum of the all possible products created from elements $X_{k}$
in such a way that each product contains element $X_{k}$ just $p_{k}-times.$
This symbol we shall call combinator.
\end{definition}

\begin{example}
\begin{equation}
\{X,Y\}=XY+YX,  \label{ga8}
\end{equation}%
\begin{equation}
\{X,Y^{2}\}=XY^{2}+YXY+Y^{2}X,  \label{ga9}
\end{equation}%
\begin{equation}
\{X,Y,Z\}=XYZ+XZY+YXZ+YZX+ZXY+ZYX.  \label{ga10}
\end{equation}
\end{example}

\noindent Now, we shall prove some useful identities.

\begin{lemma}
Let us assume $z$ is a complex variable, $p,r\geq 0$ and denote 
\begin{equation}
q_{p}(z)=(1-z)(1-z^{2})...(1-z^{p}),\qquad q_{0}(z)=1,  \label{ga11}
\end{equation}%
\begin{equation}
F_{rp}(z)=\sum_{k_{p}=0}^{r}...\sum_{k_{2}=0}^{k_{3}}%
\sum_{k_{1}=0}^{k_{2}}z^{k_{1}}z^{k_{2}}...z^{k_{p}},  \label{gaa11}
\end{equation}%
\begin{equation}
G_{p}(z)=\sum_{k=0}^{p}\frac{z^{k}}{q_{p-k}(z^{-1})q_{k}(z)},  \label{ga12}
\end{equation}%
\begin{equation}
H_{p}(z)=\sum_{k=0}^{p}\frac{1}{q_{p-k}(z^{-1})q_{k}(z)}.  \label{ga13}
\end{equation}%
Then the following identities hold for $z\neq 0,\ z^{j}\neq 1;\ j=1,2,...,p$%
: 
\begin{equation}
q_{p}(z)=(-1)^{p}z^{\frac{p(p+1)}{2}}q_{p}(z^{-1}),  \label{ga14}
\end{equation}%
\begin{equation}
G_{p}(z)=0,  \label{ga15}
\end{equation}%
\begin{equation}
H_{p}(z)=1,  \label{ga16}
\end{equation}%
\begin{equation}
F_{rp}(z)=\sum_{k=0}^{p}\frac{z^{k\cdot r}}{q_{p-k}(z)q_{k}(z^{-1})}
\label{gaa16}
\end{equation}%
and in particular for $z^{p+r}=1$%
\begin{equation}
F_{rp}(z)=0.  \label{gab16}
\end{equation}
\end{lemma}

\noindent \emph{Proof:}

\noindent 1) Relation (\ref{ga14}) follows immediately from definition (\ref%
{ga11}):%
\begin{equation*}
q_{r}(z)=(1-z)(1-z^{2})...(1-z^{r})=z\cdot
z^{2}...z^{r}(z^{-1}-1)...(z^{-r}-1)
\end{equation*}%
\begin{equation*}
=(-1)^{r}z^{\frac{r(r+1)}{2}}q_{r}(z^{-1}).
\end{equation*}%
2) Relations (\ref{ga15}), (\ref{ga16}):

\noindent First, if we invert the order of adding in the relations (\ref%
{ga12}), (\ref{ga13}) making substitution $j=p-k,$ then 
\begin{equation}
G_{p}(z)=\sum_{k=0}^{p}\frac{z^{k}}{q_{p-k}(z^{-1})q_{k}(z)}%
=z^{p}\sum_{j=0}^{p}\frac{z^{-j}}{q_{j}(z^{-1})q_{p-j}(z)}%
=z^{p}G_{p}(z^{-1}),  \label{ga17}
\end{equation}%
\begin{equation}
H_{p}(z)=\sum_{k=0}^{p}\frac{1}{q_{p-k}(z^{-1})q_{k}(z)}=\sum_{j=0}^{p}\frac{%
1}{q_{j}(z^{-1})q_{p-j}(z)}=H_{p}(z^{-1}).  \label{ga18}
\end{equation}%
Now, let us calculate%
\begin{equation}
H_{p}(z)-H_{p-1}(z)=\sum_{k=0}^{p}\frac{1}{q_{p-k}(z^{-1})q_{k}(z)}%
-\sum_{k=0}^{p-1}\frac{1}{q_{p-1-k}(z^{-1})q_{k}(z)}  \label{ga19}
\end{equation}%
\begin{equation*}
=\frac{1}{q_{p}(z)}+\sum_{k=0}^{p-1}\frac{1}{q_{p-k}(z^{-1})q_{k}(z)}%
-\sum_{k=0}^{p-1}\frac{1}{q_{p-k-1}(z^{-1})q_{k}(z)}
\end{equation*}%
\begin{equation*}
=\frac{1}{q_{p}(z)}+\sum_{k=0}^{p-1}\frac{1-(1-z^{k-p})}{%
q_{p-k}(z^{-1})q_{k}(z)}=\sum_{k=0}^{p}\frac{z^{k-p}}{q_{p-k}(z^{-1})q_{k}(z)%
}=G_{p}(z^{-1}).
\end{equation*}%
The last relation combined with Eq. (\ref{ga18}) implies 
\begin{equation}
G_{p}(z^{-1})=G_{p}(z),  \label{ga20}
\end{equation}%
which compared with Eq. (\ref{ga17}) gives 
\begin{equation}
G_{p}(z^{-1})=0;\qquad z\neq 0,\ z^{j}\neq 1,\ j=1,2,...p.  \label{ga21}
\end{equation}%
So the identity (\ref{ga15}) is proved. Further, relations (\ref{ga21}), (%
\ref{ga19}) imply 
\begin{equation}
H_{p}(z)-H_{p-1}(z)=0,  \label{ga22}
\end{equation}%
therefore 
\begin{equation}
H_{p}(z)=H_{p-1}(z)=...=H_{0}(z)=1  \label{ga23}
\end{equation}%
and the identity (\ref{ga16}) is proved as well.

\noindent 3) The relation (\ref{gaa16}) can be proved by the induction,
therefore first let us assume $p=1$, then its l.h.s. reads%
\begin{equation*}
\sum_{k_1=0}^{k_2}z^{k_1}=\frac{1-z^{k_2+1}}{1-z}
\end{equation*}
and r.h.s. gives%
\begin{equation*}
\frac 1{q_1(z)}+\frac{z^{k_2}}{q_1(z^{-1})}=\frac 1{1-z}+\frac{z^{k_2}}{%
1-z^{-1}}=\frac{1-z^{k_2+1}}{1-z},
\end{equation*}
so for $p=1$ the relation is valid. Now let us suppose the relation holds
for $p$ and calculate the case $p+1$%
\begin{equation*}
\sum_{k_{p+1}=0}^{k_{p+2}}...\sum_{k_2=0}^{k_3}%
\sum_{k_1=0}^{k_2}z^{k_1}z^{k_2}...z^{k_{p+1}}=%
\sum_{k_{p+1}=0}^{k_{p+2}}z^{k_{p+1}}...\sum_{k_2=0}^{k_3}%
\sum_{k_1=0}^{k_2}z^{k_1}z^{k_2}...z^{k_p}
\end{equation*}
\begin{equation*}
=\sum_{k_{p+1}=0}^{k_{p+2}}z^{k_{p+1}}\sum_{k=0}^p\frac{z^{k\cdot k_{p+1}}}{%
q_{p-k}(z)q_k(z^{-1})}=\sum_{k=0}^p\frac
1{q_{p-k}(z)q_k(z^{-1})}\sum_{k_{p+1}=0}^{k_{p+2}}z^{(k+1)\cdot k_{p+1}}
\end{equation*}
\begin{equation*}
=\sum_{k=0}^p\frac 1{q_{p-k}(z)q_k(z^{-1})}\frac{1-z^{(k+1)\cdot
(k_{p+2}+1)} }{1-z^{k+1}}=\sum_{k=0}^p\frac{z^{-k-1}-z^{(k+1)\cdot k_{p+2}}}{%
q_{p-k}(z)q_k(z^{-1})(z^{-k-1}-1)}
\end{equation*}
\begin{equation*}
=\sum_{k=0}^p\frac{z^{(k+1)\cdot k_{p+2}}-z^{-k-1}}{q_{p-k}(z)q_{k+1}(z^{-1})%
}=\sum_{k=1}^{p+1}\frac{z^{k\cdot k_{p+2}}-z^{-k}}{q_{p+1-k}(z)q_k(z^{-1})}%
=\sum_{k=0}^{p+1}\frac{z^{k\cdot k_{p+2}}-z^{-k}}{q_{p+1-k}(z)q_k(z^{-1})}
\end{equation*}
\begin{equation*}
=\sum_{k=0}^{p+1}\frac{z^{k\cdot k_{p+2}}}{q_{p+1-k}(z)q_k(z^{-1})}%
-\sum_{k=0}^{p+1}\frac{z^{-k}}{q_{p+1-k}(z)q_k(z^{-1})}.
\end{equation*}
The last sum equals $G_{p+1}(z^{-1}),$ which is zero according to Eq. (\ref%
{ga15}), so we have proven relation (\ref{gaa16}) for $p+1.$ Therefore the
relation is valid for any $p$.

\noindent 4) The relation (\ref{gab16}) is a special case of Eq. (\ref{gaa16}%
). The denominators in the sum (\ref{gaa16}) can be with the use of the
identity (\ref{ga14}) expressed%
\begin{equation*}
q_{p-k}(z)q_{k}(z^{-1})=(-1)^{p}z^{s}q_{p-k}(z^{-1})q_{k}(z),\qquad s=\left( 
\frac{p}{2}-k\right) (p+1)
\end{equation*}%
and since $z^{r\cdot k}=z^{-p\cdot k}$, the sum can be rewritten%
\begin{equation*}
\sum_{k=0}^{p}\frac{z^{k\cdot r}}{q_{p-k}(z)q_{k}(z^{-1})}%
=(-1)^{p}\sum_{k=0}^{p}\frac{z^{-s}z^{-p\cdot k}}{q_{p-k}(z^{-1})q_{k}(z)}
\end{equation*}%
\begin{equation*}
=(-1)^{p}z^{-\frac{p(p+1)}{2}}\sum_{k=0}^{p}\frac{z^{k}}{%
q_{p-k}(z^{-1})q_{k}(z)}.
\end{equation*}%
Obviously, the last sum coincides with $G_{p}(z)$, which is zero according
to already proven identity (\ref{ga15}).

Let us remark, last lemma implies also the known formula 
\begin{equation}
x^{n}-y^{n}=(x-y)(x-\alpha y)(x-\alpha ^{2}y)...(x-\alpha ^{n-1}y),\qquad
\alpha =\exp (2\pi i/n).  \label{gaa23}
\end{equation}%
The product can be expanded%
\begin{equation*}
x^{n}-y^{n}=\sum_{j=0}^{n}c_{j}x^{n-j}(-y)^{j}
\end{equation*}%
and one can easily check that%
\begin{equation*}
c_{0}=1,\qquad c_{n}=\alpha \alpha ^{2}\alpha ^{3}...\alpha
^{n-1}=(-1)^{n-1}.
\end{equation*}%
For the remaining $j,$ $0<j<n$ we get%
\begin{equation*}
c_{j}=\sum_{k_{j}=j-1}^{n-1}...\sum_{k_{2}=1}^{k_{3}-1}%
\sum_{k_{1}=0}^{k_{2}-1}\alpha ^{k_{1}}\alpha ^{k_{2}}...\alpha ^{k_{j}}
\end{equation*}%
and after the shift of summing limits we obtain%
\begin{equation*}
c_{j}=\alpha \alpha ^{2}\alpha ^{3}...\alpha
^{j-1}\sum_{k_{j}=0}^{n-j}...\sum_{k_{2}=0}^{k_{3}}\sum_{k_{1}=0}^{k_{2}}%
\alpha ^{k_{1}}\alpha ^{k_{2}}...\alpha ^{k_{j}}.
\end{equation*}%
This multiple sum is a special case of the formula (\ref{gaa11}) and since $%
\alpha ^{n}=1$, the identity (\ref{gab16}) is satisfied. Therefore for $0<j<n
$ we get $c_{j}=0$ and formula (\ref{gaa23}) is proved.

\begin{definition}
Let us have a matrix product created from some string of matrices $X,Y$ in
such a way that matrix $X$ is in total involved $p-times$ and $Y\quad
r-times $. By the symbol $P_j^{+}$ ($P_j^{-}$) we denote permutation, which
shifts the leftmost (rightmost) matrix to right (left) on the position in
which the shifted matrix has $j$ matrices of different kind left (right).
(Range of $j$ is restricted by $p$ or $r$ if the shifted matrix is $Y$ or $X$%
).
\end{definition}

\begin{example}
\begin{equation}  \label{ga24}
P_3^{+}\circ XYXYYXY=YXYYXXY
\end{equation}
\end{example}

\noindent Now, we can prove the following theorem.

\begin{theorem}
\label{tst}Let $p,r>0$ and $p+r=n$ (i.e. $\alpha ^{p+r}=1)$. Then the
matrices $S,T$ fulfill 
\begin{equation}  \label{ga25}
\{S^p,T^r\}=0.
\end{equation}
\end{theorem}

\noindent \textit{Proof:}

\noindent Obviously, all the terms in the combinator $\{S^{p},T^{r}\}$ can
be generated e.g. from the string 
\begin{equation*}
\underset{p}{{\underbrace{SS....S}}}{\ }\underset{r}{{\underbrace{TT....T}}}%
=S^{p}T^{r}
\end{equation*}%
by means of the permutations $P_{j}^{+}$%
\begin{equation}
\{S^{p},T^{r}\}=\sum_{k_{p}=0}^{r}...\sum_{k_{2}=0}^{k_{3}}%
\sum_{k_{1}=0}^{k_{2}}P_{k_{1}}^{+}\circ P_{k_{2}}^{+}...P_{k_{p}}^{+}\circ
S^{p}T^{r}.  \label{ga26}
\end{equation}%
Now the relation (\ref{ga2}) implies%
\begin{equation*}
P_{j}^{+}\circ S^{p}T^{r}=\alpha ^{j}S^{p}T^{r}
\end{equation*}%
and Eq. (\ref{ga26}) can be modified 
\begin{equation}
\{S^{p},T^{r}\}=\left(
\sum_{k_{p}=0}^{r}...\sum_{k_{2}=0}^{k_{3}}\sum_{k_{1}=0}^{k_{2}}\alpha
^{k_{1}}\alpha ^{k_{2}}...\alpha ^{k_{p}}\right) S^{p}T^{r}.  \label{gaa26}
\end{equation}%
Apparently the multiple sum in this equation coincides with r.h.s. of Eq. (%
\ref{gaa11}) and satisfies the condition for Eq. (\ref{gab16}), thereby the
theorem is proved.

Let us remark, that alternative use of permutations $P_{j}^{-}$ instead of $%
P_{j}^{+}$ would lead to the equation 
\begin{equation}
\{S^{p},T^{r}\}=\left(
\sum_{k_{r}=0}^{p}...\sum_{k_{2}=0}^{k_{3}}\sum_{k_{1}=0}^{k_{2}}\alpha
^{k_{1}}\alpha ^{k_{2}}...\alpha ^{k_{r}}\right) S^{p}T^{r}.  \label{gab26}
\end{equation}%
Comparison of Eqs. (\ref{gaa26}), (\ref{gab26}) with the relation for $F_{pr}
$ defined by Eq. (\ref{gaa11}) implies 
\begin{equation}
F_{pr}(\alpha )=F_{rp}(\alpha ).  \label{gac26}
\end{equation}%
Obviously this equation is valid irrespective of the assumption $\alpha
^{p+r}=1$, i.e. it holds for any $n$ and $\alpha =\exp (2\pi i/n)$. It
follows, that Eq. (\ref{gac26}) is satisfied for any $\alpha $.

\begin{definition}
\label{def9}By the symbols $Q_{pr}$ we denote $n^{2}$ matrices 
\begin{equation}
Q_{pr}=S^{p}T^{r},\qquad p,r=1,2,...,n  \label{ga27}
\end{equation}
\end{definition}

\begin{lemma}
\begin{equation}  \label{gaa27}
Q_{rs}Q_{pq}=\alpha ^{s\cdot p}Q_{kl};\ k=\text{mod}(r+p-1,n)+1,\ l=\text{mod%
}(s+q-1,n)+1,
\end{equation}
\begin{equation}  \label{ga28}
Q_{rs}Q_{pq}=\alpha ^{s\cdot p-r\cdot q}Q_{pq}Q_{rs},
\end{equation}
\begin{equation}  \label{ga29}
\left( Q_{rs}\right) ^n=(-1)^{(n-1)r\cdot s}I,
\end{equation}
\begin{equation}  \label{ga30}
Q_{rs}^{\dagger }Q_{rs}=Q_{rs}Q_{rs}^{\dagger }=I,
\end{equation}
\begin{equation}  \label{gaa30}
Q_{rs}^{\dagger }=\alpha ^{r\cdot s}Q_{kl};\qquad k=n-r,\qquad l=n-s,
\end{equation}
\begin{equation}  \label{ga31}
\det Q_{rs}=(-1)^{(n-1)(r+s)}
\end{equation}
and for $r\neq n$ or $s\neq n$ 
\begin{equation}  \label{ga32}
\text{\textrm{Tr }}Q_{rs}=0.
\end{equation}
\end{lemma}

\noindent \textit{Proof:}

\noindent The relations follow from the definition of $Q_{pr}$ and relations
(\ref{ga2})-(\ref{ga6}).

\begin{theorem}
The matrices $Q_{pr}$ are linearly independent and any matrix $A$ (of the
same dimension) can be expressed as their linear combination 
\begin{equation}
A=\sum_{k,l=1}^{n}a_{kl}Q_{kl},\qquad a_{kl}=\frac{1}{n}\text{\textrm{Tr}}%
(Q_{kl}^{\dagger }A).  \label{ga33}
\end{equation}
\end{theorem}

\noindent \textit{Proof:}

\noindent Let us assume matrices $Q_{kl}$ are linearly dependent, i.e. there
exists some $a_{rs}\neq 0$ and simultaneously%
\begin{equation*}
\sum_{k,l=1}^{n}a_{kl}Q_{kl}=0,
\end{equation*}%
which with the use of the previous lemma implies%
\begin{equation*}
\text{\textrm{Tr}}\sum_{k,l=1}^{n}a_{kl}Q_{rs}^{\dagger }Q_{kl}=a_{rs}n=0.
\end{equation*}%
This equation contradicts our assumption, therefore the matrices are
independent and obviously represent a base in the linear space of matrices $%
n\times n$, which with the use of the previous lemma implies the relations (%
\ref{ga33}).

\begin{theorem}
For any $n\geq 2$, among $n^2$ matrices (\ref{ga27}) there exists the triad $%
Q_\lambda ,Q_\mu ,Q_\nu $ for which 
\begin{equation}  \label{ga34}
\{Q_\lambda ^p,Q_\mu ^r\}=\{Q_\mu ^p,Q_\nu ^r\}=\{Q_\nu ^p,Q_\lambda
^r\}=0;\qquad 0<p,r,\qquad p+r=n
\end{equation}
and moreover if $n\geq 3$, then also 
\begin{equation}  \label{ga35}
\{Q_\lambda ^p,Q_\mu ^r,Q_\nu ^s\}=0;\qquad 0<p,r,s,\qquad p+r+s=n.
\end{equation}
\end{theorem}

\noindent \textit{Proof:}

\noindent We shall show the relations hold e.g. for indices $\lambda =1n,\
\mu =11,\ \nu =n1.$ Let us denote 
\begin{equation}
X=Q_{1n}=S,\qquad Y=Q_{11},\qquad Z=Q_{n1}=T,  \label{ga35.1}
\end{equation}%
then the relation (\ref{ga28}) implies 
\begin{equation}
YX=\alpha XY,\qquad ZX=\alpha XZ,\qquad ZY=\alpha YZ.  \label{gaa35}
\end{equation}%
Actually the relation $\{X^{p},Z^{r}\}=0$ is already proven in the Theorem %
\ref{tst}, obviously the remaining relations (\ref{ga34}) can be proved
exactly in the same way.

The combinator (\ref{ga35}) can be similarly as in the proof of Theorem \ref%
{tst} expressed 
\begin{equation}
\{X^{p},Y^{r},Z^{s}\}  \label{ga36}
\end{equation}%
\begin{equation*}
=\sum_{j_{p}=0}^{r+s}...\sum_{j_{2}=0}^{j_{3}}%
\sum_{j_{1}=0}^{j_{2}}P_{j_{1}}^{+}\circ P_{j_{2}}^{+}...P_{j_{p}}^{+}\circ
X^{p}\sum_{k_{p}=0}^{s}...\sum_{k_{2}=0}^{k_{3}}%
\sum_{k_{1}=0}^{k_{2}}P_{k_{1}}^{+}\circ P_{k_{2}}^{+}...P_{k_{r}}^{+}\circ
Y^{r}Z^{s},
\end{equation*}%
which for matrices obeying relations (\ref{gaa35}) give%
\begin{equation*}
\{X^{p},Y^{r},Z^{s}\}
\end{equation*}%
\begin{equation*}
=\left(
\sum_{j_{p}=0}^{r+s}...\sum_{j_{2}=0}^{j_{3}}\sum_{j_{1}=0}^{j_{2}}\alpha
^{j_{1}}\alpha ^{j_{2}}...\alpha ^{j_{p}}\right) \left(
\sum_{k_{p}=0}^{s}...\sum_{k_{2}=0}^{k_{3}}\sum_{k_{1}=0}^{k_{2}}\alpha
^{k_{1}}\alpha ^{k_{2}}...\alpha ^{k_{r}}\right) X^{p}Y^{r}Z^{s}.
\end{equation*}%
Since the first multiple sum (with indices $j$) coincides with Eq. (\ref%
{gaa11}) and satisfy the condition for Eq. (\ref{gab16}), r.h.s. is zero and
the theorem is proved.

Now let us make few remarks to illuminate content of the last theorem and
meaning of the matrices $Q_{\lambda }$.\textit{\ }Obviously, the relations (%
\ref{ga34}), (\ref{ga35}) are equivalent to the statement: any three complex
numbers $a,b,c$ satisfy 
\begin{equation}
(aQ_{\lambda }+bQ_{\mu }+cQ_{\nu })^{n}=(a^{n}+b^{n}+c^{n})I.  \label{ga37}
\end{equation}%
Further, the theorem speaks about existence of the triad but not about their
number. Generally for $n>2$ there is more than one triad defined by the
theorem, but on the other hand not any three various matrices from the set $%
Q_{rs}$ comply with the theorem. Simple example are some $X,Y,Z$ where e.g. $%
XY=YX,$ which happens for $Y\sim X^{p},2\leq p<n$. Obviously in this case at
least the relation (\ref{ga34}) surely is not satisfied. Computer check of
the relation (\ref{ga36}) which has been done with all possible triads from $%
Q_{rs}$ for $2\leq n\leq 20$ suggests, that a triad $X,Y,Z$ for which there
exist the numbers $p,r,s\geq 1$ and $p+r+s\leq n$ so that $%
X^{p}Y^{r}Z^{s}\sim I$ also does not comply with the theorem. Further, the
result on r.h.s. of Eq. (\ref{ga36}) generally depends on the factors $\beta
_{k}$ in relations 
\begin{equation}
XY=\beta _{3}YX\qquad YZ=\beta _{1}ZY\qquad ZX=\beta _{2}XZ  \label{gaa37}
\end{equation}%
and computer check suggests the sets in which for some $\beta _{k}$ and $p<n$
there is $\beta _{k}^{p}=1$ also contradict the theorem. In this way the
number of different triads obeying the relations (\ref{ga34}), (\ref{ga35})
is rather complicated function of $n$ - as shown in the table 
\begin{equation*}
\begin{array}{cccccccccccccccccccc}
n: & \text{{\footnotesize 2}} & \text{{\footnotesize 3}} & \text{%
{\footnotesize 4}} & \text{{\footnotesize 5}} & \text{{\footnotesize 6}} & 
\text{{\footnotesize 7}} & \text{{\footnotesize 8}} & \text{{\footnotesize 9}%
} & \text{{\footnotesize 10}} & \text{{\footnotesize 11}} & \text{%
{\footnotesize 12}} & \text{{\footnotesize 13}} & \text{{\footnotesize 14}}
& \text{{\footnotesize 15}} & \text{{\footnotesize 16}} & \text{%
{\footnotesize 17}} & \text{{\footnotesize 18}} & \text{{\footnotesize 19}}
& \text{{\footnotesize 20}} \\ 
\#3: & \text{{\footnotesize 1}} & \text{{\footnotesize 1}} & \text{%
{\footnotesize 1}} & \text{{\footnotesize 4}} & \text{{\footnotesize 1}} & 
\text{{\footnotesize 9}} & \text{{\footnotesize 4}} & \text{{\footnotesize 9}%
} & \text{{\footnotesize 4}} & \text{{\footnotesize 25}} & \text{%
{\footnotesize 4}} & \text{{\footnotesize 36}} & \text{{\footnotesize 9}} & 
\text{{\footnotesize 16}} & \text{{\footnotesize 16}} & \text{{\footnotesize %
64}} & \text{{\footnotesize 9}} & \text{{\footnotesize 81}} & \text{%
{\footnotesize 16}}%
\end{array}%
\end{equation*}%
Here the statement triad $X,Y,Z$ is different from $X^{\prime },Y^{\prime
},Z^{\prime }$ means that after any rearrangement of the symbols $X,Y,Z$ for
marking of matrices in the given set, there is always at least one pair $%
\beta _{k}\neq \beta _{k}^{\prime }.$

\textit{\ }Naturally, one can ask if there exists also the set of four or
generally $N$ matrices, which satisfy a relation similar to Eq. (\ref{ga37}) 
\begin{equation}
\left( \sum_{\lambda =0}^{N-1}a_{\lambda }Q_{\lambda }\right)
^{n}=\sum_{\lambda =0}^{N-1}a_{\lambda }^{n}.  \label{ga38}
\end{equation}%
For $2\leq n\leq 10$ and $N=4$ the computer suggests the negative answer -
in the case of matrices generated according to the \textit{Definition \ref%
{def9}.} However, one can verify: if $U_{l},l=1,2,3$ is the triad complying
with the theorem (or equivalently with the relation (\ref{ga37})), then the
matrices $n^{2}\times n^{2}$%
\begin{equation}
Q_{0}=I\otimes T=\left( 
\begin{array}{cccccc}
I & \qquad  & \qquad  & \qquad  & \qquad  &  \\ 
\qquad  & \alpha I &  &  &  &  \\ 
&  & \alpha ^{2}I &  &  &  \\ 
&  &  & \cdot  &  &  \\ 
&  &  &  & \cdot  &  \\ 
&  &  &  &  & \alpha ^{n-1}I%
\end{array}%
\right) ,  \label{ga39}
\end{equation}%
\begin{equation}
Q_{l}=U_{l}\otimes S=\left( 
\begin{array}{cccccc}
0 & \qquad  & \qquad  & \qquad  & \qquad  & U_{l} \\ 
U_{l} &  &  &  &  & \qquad  \\ 
\qquad  & U_{l} &  &  &  &  \\ 
&  & \cdot  &  &  &  \\ 
&  &  & \cdot  &  &  \\ 
&  &  &  & U_{l} & 0%
\end{array}%
\right)   \label{gb39}
\end{equation}%
satisfy relation (\ref{ga38}) for $N=4$. Generally, if $U_{\lambda }$ are
matrices complying with Eq. (\ref{ga38}) for some $N\geq 3$, then the
matrices created from them according to the rule (\ref{ga39}), (\ref{gb39})
will satisfy Eq. (\ref{ga38}) for $N+1$. The last statement follows from the
following equalities. Let us assume%
\begin{equation*}
\sum_{k=0}^{N}p_{k}=n,
\end{equation*}%
then%
\begin{equation*}
\{Q_{0}^{p_{0}},Q_{1}^{p_{1}},...Q_{N}^{p_{N}}\}=%
\sum_{j_{p_{N}}=0}^{n-p_{N}}...\sum_{j_{1}=0}^{j_{2}}%
\sum_{j_{0}=0}^{j_{1}}P_{j_{0}}^{-}\circ
P_{j_{1}}^{-}...P_{j_{p_{N}}}^{-}\circ
\{Q_{0}^{p_{0}},...Q_{N-1}^{p_{N-1}}\}Q_{N}^{p_{N}}
\end{equation*}%
\begin{equation*}
=\sum_{j_{p_{N}}=0}^{n-p_{N}}...\sum_{j_{1}=0}^{j_{2}}%
\sum_{j_{0}=0}^{j_{1}}P_{j_{0}}^{-}\circ
P_{j_{1}}^{-}...P_{j_{p_{N}}}^{-}\circ \{(U_{0}\otimes
S)^{p_{0}},...(U_{N-1}\otimes S)^{p_{N-1}}\}(I\otimes T)^{p_{N}}
\end{equation*}%
\begin{equation*}
=\sum_{j_{p_{N}}=0}^{n-p_{N}}...\sum_{j_{1}=0}^{j_{2}}\sum_{j_{0}=0}^{j_{1}}%
\alpha ^{j_{0}}\alpha ^{j_{1}}...\alpha ^{j_{p_{N}}}\{(U_{0}\otimes
S)^{p_{0}},...(U_{N-1}\otimes S)^{p_{N}-1}\}(I\otimes T)^{p_{N}}
\end{equation*}%
\begin{equation*}
=\left(
\sum_{j_{p_{N}}=0}^{n-p_{N}}...\sum_{j_{1}=0}^{j_{2}}\sum_{j_{0}=0}^{j_{1}}%
\alpha ^{j_{0}}\alpha ^{j_{1}}...\alpha ^{j_{p_{N}}}\right)
\{U_{1}^{p_{1}},...U_{N-1}^{p_{N-1}}\}\otimes S^{n-p_{N}}T^{p_{N}},
\end{equation*}%
where the last multiple sum equals zero according to the relations (\ref%
{gaa11}) and (\ref{gab16}). Obviously for $n=2$ the matrices (\ref{ga35.1})
and (\ref{ga39}),(\ref{gb39}) created from them correspond, up to some phase
factors, to the Pauli matrices $\sigma _{j}$ and Dirac matrices $\gamma
_{\mu }$.

Obviously, from the set of matrices $Q_{rs}$ (with exception of $Q_{nn}=I$)
one can easily make the $n^{2}-1$ generators of the fundamental
representation of $SU(n)$ group 
\begin{equation}
G_{rs}=a_{rs}Q_{rs}+a_{rs}^{\ast }Q_{rs}^{+},  \label{ga40}
\end{equation}%
where $a_{rs}$ are suitable factors. For example the choice 
\begin{equation}
a_{kl}=\frac{1}{\sqrt{2}}\alpha ^{\lbrack kl+n(k+l-1/4)]/2}  \label{ga41}
\end{equation}%
gives commutation relations%
\begin{equation}
\left[ G_{kl},G_{rs}\right] =i\sin \left( \pi (ks-lr)/n\right)   \label{ga42}
\end{equation}%
\begin{equation*}
\cdot \left\{ \mathrm{sg}(k+r,l+s,n)\left(
G_{k+r,l+s}-(-1)^{n+k+l+r+s}G_{-k-r,-l-s}\right) \right. 
\end{equation*}%
\begin{equation*}
-\left. \mathrm{sg}(k-r,l-s,n)\left(
G_{k-r,l-s}-(-1)^{n+k+l+r+s}G_{r-k,s-l}\right) \right\} ,
\end{equation*}%
where%
\begin{equation*}
\mathrm{sg}(p,q,n)=(-1)^{p\cdot m_{q}+q\cdot m_{p}-n},\qquad m_{x}=\frac{x-%
\text{mod}(x-1,n)-1}{n}.
\end{equation*}%
and indices at $G$ (on r.h.s.) in Eq. (\ref{ga42}) are understood in the
sense of mod - like in the relation (\ref{gaa27}). One can easily check e.g.
for $n=2$ the matrices (\ref{ga40}) with the factors $a_{rs}$ according to
Eq. (\ref{ga41}) are the Pauli matrices - generators of the fundamental
representation of the $SU(2)$ group.

\section{Wave equations generated by the roots \newline
of D'Alambertian operator $\square ^{1/n}$}

Now, using the generalized Dirac matrices (\ref{ga39}), (\ref{gb39}) we
shall assemble the corresponding wave equation as follows. These four
matrices with normalization 
\begin{equation}
\left( Q_{0}\right) ^{n}=-\left( Q_{l}\right) ^{n}=I,\qquad l=1,2,3,
\label{s1}
\end{equation}%
allow to write down the set of algebraic equations 
\begin{equation}
\left( \Gamma (p)-\mu I\right) \Psi (p)=0,  \label{s2}
\end{equation}%
where%
\begin{equation}
\Gamma (p)=\sum_{\lambda =0}^{3}\pi _{\lambda }Q_{\lambda }.  \label{s3}
\end{equation}%
If the variables $\mu ,\ \pi _{\lambda }$ represent the fractional powers of
the mass and momentum components%
\begin{equation}
\mu ^{n}=m^{2},\qquad \pi _{\lambda }^{n}=p_{\lambda }^{2},  \label{sa3}
\end{equation}%
then 
\begin{equation}
\Gamma (p)^{n}=p_{0}^{2}-p_{1}^{2}-p_{2}^{2}-p_{3}^{2}\equiv p^{2}
\label{sb3}
\end{equation}%
and after $n-1$ times repeated application of the operator $\Gamma $ on Eq. (%
\ref{s2}) one gets the set of Klein-Gordon equations in the $p-$%
representation 
\begin{equation}
\left( p^{2}-m^{2}\right) \Psi (p)=0.  \label{s4}
\end{equation}%
The Eqs. (\ref{s2}) and (\ref{s4}) are the sets of $n^{2}$ equations with
solution $\Psi $ having $n^{2}$ components. Obviously, the case $n^{2}=4$
corresponds to the Dirac equation. For $n>2$ the Eq. (\ref{s2}) is new, more
complicated and immediately invoking some questions. In the present paper we
shall attempt to answer at least some of them. One can check, that the
solution of the set (\ref{s2}) reads%
\begin{equation}
\Psi (p)=\left( 
\begin{array}{c}
\mathbf{h} \\ 
\frac{U(p)}{\alpha \pi _{0}-\mu }\mathbf{h} \\ 
\frac{U^{2}(p)}{(\alpha \pi _{0}-\mu )(\alpha ^{2}\pi _{0}-\mu )}\mathbf{h}
\\ 
\cdot  \\ 
\cdot  \\ 
\frac{U^{n-1}(p)}{(\alpha \pi _{0}-\mu )...(\alpha ^{n-1}\pi _{0}-\mu )}%
\mathbf{h}%
\end{array}%
\right) ,\qquad \mathbf{h}=\left( 
\begin{array}{c}
h_{1} \\ 
h_{2} \\ 
\cdot  \\ 
\cdot  \\ 
h_{n}%
\end{array}%
\right) ,  \label{sa4}
\end{equation}%
where 
\begin{equation*}
U(p)=\sum_{l=1}^{3}\pi _{l}U_{l},\qquad \left( U_{l}\right) ^{n}=-I,
\end{equation*}%
($U_{l}$ is the triad from which the matrices $Q_{l}$ are constructed in
accordance with Eqs. (\ref{ga39}), (\ref{gb39})) and $h_{1},h_{2},...h_{n}$
are arbitrary functions of $p$. At the same time, $\pi _{\lambda }$ satisfy
the constraint 
\begin{equation}
\pi _{0}^{n}-\pi _{1}^{n}-\pi _{2}^{n}-\pi _{3}^{n}=\mu ^{n}=m^{2}.
\label{sb4}
\end{equation}

First of all, one can bring to notice, that in Eq. (\ref{s2}) the fractional
powers of the momentum components appear, which means that the equation in
the $x-$representation will contain the fractional derivatives: 
\begin{equation}
\pi _{\lambda }=(p_{\lambda })^{2/n}\rightarrow (i\partial _{\lambda
})^{2/n}.  \label{s5}
\end{equation}%
Our primary considerations will concern $p-$representation, but afterwards
we shall show how the transition to the $x-$representation can be realized
by means of the Fourier transformation, in accordance with the approach
suggested in \cite{zav}.

Further question concerns relativistic covariance of Eq. (\ref{s2}): How to
transform simultaneously the operator 
\begin{equation}
\Gamma (p)\rightarrow \Gamma (p^{\prime })=\Lambda \Gamma (p)\Lambda ^{-1}
\label{s6}
\end{equation}%
and the solution 
\begin{equation}
\Psi (p)\rightarrow \Psi ^{\prime }(p^{\prime })=\Lambda \Psi (p)  \label{s7}
\end{equation}%
to preserve the equal form of the operator $\Gamma $ for initial variables $%
p_{\lambda }$ and the boosted ones $p_{\lambda }^{\prime }$ ?

\subsection{Infinitesimal transformations}

First let us consider the infinitesimal transformations 
\begin{equation}
\Lambda (d\omega )=I+id\omega \cdot L_{\omega },  \label{s8}
\end{equation}%
where $d\omega $ represents the infinitesimal values of the six parameters
of the Lorentz group corresponding to the space rotations 
\begin{equation}
p_{i}^{\prime }=p_{i}+\epsilon _{ijk}p_{j}d\varphi _{k},\qquad i=1,2,3
\label{s9}
\end{equation}%
and the Lorentz transformations 
\begin{equation}
p_{i}^{\prime }=p_{i}+p_{0}d\psi _{i},\qquad p_{0}^{\prime
}=p_{0}+p_{i}d\psi _{i},\qquad i=1,2,3,  \label{s10}
\end{equation}%
where $\tanh \psi _{i}=v_{i}/c\equiv \beta _{i}$ is the corresponding
velocity. Here, and anywhere in the next we use the convention, that in the
expressions involving the antisymmetric tensor $\epsilon _{ijk}$, the
summation over indices appearing twice is done. From the infinitesimal
transformations (\ref{s9}), (\ref{s10}) one can obtain the finite ones. For
the three space rotations we get 
\begin{equation}
p_{1}^{\prime }=p_{1}\cos \varphi _{3}+p_{2}\sin \varphi _{3},\qquad
p_{2}^{\prime }=p_{2}\cos \varphi _{3}-p_{1}\sin \varphi _{3},\qquad
p_{3}^{\prime }=p_{3},  \label{sa10}
\end{equation}%
\begin{equation}
p_{2}^{\prime }=p_{2}\cos \varphi _{1}+p_{3}\sin \varphi _{1},\qquad
p_{3}^{\prime }=p_{3}\cos \varphi _{1}-p_{2}\sin \varphi _{1},\qquad
p_{1}^{\prime }=p_{1},  \label{sb10}
\end{equation}%
\begin{equation}
p_{3}^{\prime }=p_{3}\cos \varphi _{2}+p_{1}\sin \varphi _{2},\qquad
p_{1}^{\prime }=p_{1}\cos \varphi _{2}-p_{3}\sin \varphi _{2},\qquad
p_{2}^{\prime }=p_{2}  \label{sc10}
\end{equation}%
and for the Lorentz transformations similarly 
\begin{equation}
p_{0}^{\prime }=p_{0}\cosh \psi _{i}+p_{i}\sinh \psi _{i},\qquad i=1,2,3,
\label{sd10}
\end{equation}%
where 
\begin{equation}
\cosh \psi _{i}=\frac{1}{\sqrt{1-\beta _{i}^{2}}},\qquad \sinh \psi _{i}=%
\frac{\beta _{i}}{\sqrt{1-\beta _{i}^{2}}}.  \label{se10}
\end{equation}%
The definition of the six parameters implies that the corresponding
infinitesimal transformations of the reference frame $p\rightarrow p^{\prime
}$ changes a function $f(p)$ : 
\begin{equation}
f(p)\rightarrow f(p^{\prime })=f(p+\delta p)=f(p)+\frac{df}{d\omega }d\omega
,  \label{s11}
\end{equation}%
where $d/d\omega $ stands for 
\begin{equation}
\frac{d}{d\varphi _{i}}=-\epsilon _{ijk}p_{j}\frac{\partial }{\partial p_{k}}%
,\qquad \frac{d}{d\psi _{i}}=p_{0}\frac{\partial }{\partial p_{i}}+p_{i}%
\frac{\partial }{\partial p_{0}},\qquad i=1,2,3.  \label{s12}
\end{equation}%
Obviously, the equation 
\begin{equation}
p^{\prime }=p+\frac{dp}{d\omega }d\omega  \label{s13}
\end{equation}%
combined with Eq. (\ref{s12}) is identical to Eqs. (\ref{s9}), (\ref{s10}).
Further, with the use of formulas (\ref{s8}) and (\ref{s12}) the relations (%
\ref{s6}), (\ref{s7}) can be rewritten in the infinitesimal form 
\begin{equation}
\Gamma (p^{\prime })=\Gamma (p)+\frac{d\Gamma (p)}{d\omega }d\omega =\left(
I+id\omega \cdot L_{\omega }\right) \Gamma (p)\left( I-id\omega \cdot
L_{\omega }\right) ,  \label{s14}
\end{equation}%
\begin{equation}
\Psi ^{\prime }(p^{\prime })=\Psi ^{\prime }(p)+\frac{d\Psi ^{\prime }(p)}{%
d\omega }d\omega =\left( I+id\omega \cdot L_{\omega }\right) \Psi (p).
\label{s15}
\end{equation}%
If we define 
\begin{equation}
\mathbf{L_{\omega }}=L_{\omega }+i\frac{d}{d\omega },  \label{s16}
\end{equation}%
then the relations (\ref{s14}), (\ref{s15}) imply 
\begin{equation}
\lbrack \mathbf{L_{\omega }},\Gamma ]=0,  \label{s17}
\end{equation}%
\begin{equation}
\Psi ^{\prime }(p)=\left( I+id\omega \cdot \mathbf{L_{\omega }}\right) \Psi
(p).  \label{s18}
\end{equation}%
The six operators $\mathbf{L_{\omega }}$ are generators of the corresponding
representation of the Lorentz group, so they have to satisfy the commutation
relations 
\begin{equation}
\lbrack \mathbf{L_{\varphi _{j}}},\mathbf{L_{\varphi _{k}}}]=i\epsilon _{jkl}%
\mathbf{L_{\varphi _{l}},}  \label{s19}
\end{equation}%
\begin{equation}
\lbrack \mathbf{L_{\psi _{j}}},\mathbf{L_{\psi _{k}}}]=-i\epsilon _{jkl}%
\mathbf{L_{\varphi _{l}},}  \label{s20}
\end{equation}%
\begin{equation}
\lbrack \mathbf{L_{\varphi _{j}}},\mathbf{L_{\psi _{k}}}]=i\epsilon _{jkl}%
\mathbf{L_{\psi _{l}},\qquad }j,k,l=1,2,3.  \label{s21}
\end{equation}%
How this representation looks like, in other words, what operators $\mathbf{%
L_{\omega }}$ satisfy Eqs.(\ref{s19}) - (\ref{s21}) and (\ref{s17})? First,
one can easily check, that for $n>2$ there do not exist matrices $L_{\omega
} $ with constant elements representing the first term in r.h.s. of equality
(\ref{s16}) and satisfying the Eq. (\ref{s17}). If one assumes, that $%
L_{\omega }$ consist only of constant elements, then the elements of matrix $%
\frac{d}{d\omega }\Gamma (p)$ involving the terms like $p_{i}^{2/n-1}p_{j}$
certainly cannot be expressed through the elements of the difference $%
L_{\omega }\Gamma -\Gamma L_{\omega }$ consisting only of the elements
proportional to $p_{k}^{2/n}$ - in contradistinction to the case $n=2,$ i.e.
the case of Dirac equation. In this way the Eq. (\ref{s17}) cannot be
satisfied for $n>2$ and $L_{\omega }$ constant. Nevertheless, one can show,
that the set of Eqs. (\ref{s17}),(\ref{s19}) - (\ref{s21}) is solvable,
provided that we accept the elements of the matrices $L_{\omega }$ are not
constants, but the functions of $p_{i}$. To prove this, let us first make a
few preparing steps.

\begin{definition}
Let $\Gamma _{1}(p),\Gamma _{2}(p)$ and $X$ be the square matrices of the
same dimension and%
\begin{equation*}
\Gamma _{1}(p)^{n}=\Gamma _{2}(p)^{n}=p^{2}.
\end{equation*}
Then for any matrix $X$ we define the form 
\begin{equation}
Z(\Gamma _{1},X,\Gamma _{2})=\frac{1}{np^{2}}\sum_{j=1}^{n}\Gamma
_{1}^{j}X\Gamma _{2}^{n-j}.  \label{s22}
\end{equation}
\end{definition}

\noindent One can easily check, that the matrix $Z$ satisfies e.g. 
\begin{equation}
\Gamma _{1}Z=Z\Gamma _{2},  \label{s23}
\end{equation}%
\begin{equation}
Z(Z(X))=Z(X)  \label{s24}
\end{equation}%
and in particular for $\Gamma _{1}=\Gamma _{2}\equiv \Gamma $%
\begin{equation}
\lbrack \Gamma ,Z]=0,  \label{s25}
\end{equation}%
\begin{equation}
\lbrack \Gamma ,X]=0\Rightarrow X=Z(X).  \label{sa25}
\end{equation}

\begin{lemma}
\label{le4} The Eq. (\ref{s2}) can be expressed in the diagonalized
(canonical) form 
\begin{equation}
\left( \Gamma _{0}(p)-\mu \right) \Psi _{0}(p)=0;\mathbf{\qquad }\Gamma
_{0}(p)\equiv \left( p^{2}\right) ^{1/n}Q_{0,}  \label{s26}
\end{equation}%
where $Q_{0}$ is the matrix (\ref{ga39}), i.e. there exists the set of
transformations $Y$, that 
\begin{equation}
\Gamma _{0}(p)=Y(p)\Gamma (p)Y^{-1}(p);\qquad Y=Z(\Gamma _{0},X,\Gamma )
\label{s27}
\end{equation}%
and a particular form reads 
\begin{equation}
Y=y\cdot Z(\Gamma _{0},I,\Gamma ),\mathbf{\qquad }Y^{-1}=y\cdot Z(\Gamma
,I,\Gamma _{0}),  \label{sa27}
\end{equation}%
where%
\begin{equation*}
y=\sqrt{\frac{n\left[ 1-\left( p_{0}^{2}/p^{2}\right) ^{1/n}\right] }{%
1-p_{0}^{2}/p^{2}}}.
\end{equation*}
\end{lemma}

\noindent \textit{Proof:}

\noindent The Eq. (\ref{s23}) implies 
\begin{equation}
\Gamma _{0}=Z(\Gamma _{0},X,\Gamma )\Gamma Z(\Gamma _{0},X,\Gamma )^{-1},
\label{s28}
\end{equation}%
therefore, if the matrix $X$ is chosen in such a way that $\det Z\neq 0$,
then $Z^{-1}$ exists and the transformation (\ref{s28}) diagonalizes the
matrix $\Gamma $. Let us put $X=I$ and calculate the following product 
\begin{equation}
C=Z(\Gamma _{0},I,\Gamma )Z(\Gamma ,I,\Gamma _{0})=\frac{1}{n^{2}p^{4}}%
\sum_{i,j=1}^{n}\Gamma _{0}^{i}\Gamma ^{n-i+j}\Gamma _{0}^{n-j}.  \label{s29}
\end{equation}%
The last sum can be rearranged, instead of the summation index $j$ we use
the new one 
\begin{equation*}
k=i-j\mathbf{\quad }\mathrm{for}\mathbf{\ }\text{ }i\geq j,\mathbf{\quad }%
k=i-j+n\mathbf{\quad }\mathrm{for}\mathbf{\ }\text{ }i<j;\mathbf{\qquad }%
k=0,...n-1,
\end{equation*}%
then the Eq. (\ref{s29}) reads 
\begin{equation}
C=\frac{1}{n^{2}p^{4}}\sum_{k=0}^{n-1}\left( \sum_{i=k+1}^{n}\Gamma
_{0}^{i}\Gamma ^{n-k}\Gamma _{0}^{n+k-i}+\sum_{i=1}^{k}\Gamma _{0}^{i}\Gamma
^{2n-k}\Gamma _{0}^{k-i}\right)  \label{s31}
\end{equation}%
and if we take into account that $\Gamma _{0}^{n}=\Gamma ^{n}=p^{2},$ then
this sum can be simplified 
\begin{equation}
C=\sum_{k=0}^{n-1}C_{k}=\frac{1}{n^{2}p^{2}}\sum_{k=0}^{n-1}\sum_{i=1}^{n}%
\Gamma _{0}^{i}\Gamma ^{n-k}\Gamma _{0}^{k-i}.  \label{s32}
\end{equation}%
For the term $k=0$ we get 
\begin{equation}
C_{0}=\frac{1}{n^{2}p^{2}}\sum_{i=1}^{n}\Gamma _{0}^{i}\Gamma ^{n}\Gamma
_{0}^{-i}=\frac{1}{n}  \label{s33}
\end{equation}%
and for $k>0$, using Eqs. (\ref{s3}), (\ref{ga39}), (\ref{gb39}), (\ref{s26}%
) and \textit{Definition \ref{def2}} one obtains%
\begin{equation}
C_{k}=\frac{1}{n^{2}p^{2}}\sum_{i=1}^{n}\Gamma _{0}^{i}\Gamma ^{n-k}\Gamma
_{0}^{k-i}=\frac{1}{n^{2}p^{2}}\sum_{i=1}^{n}\Gamma _{0}^{i}\left(
\sum_{\lambda =0}^{3}\pi _{\lambda }Q_{\lambda }\right) ^{n-k}\Gamma
_{0}^{k-i}  \label{s34}
\end{equation}%
\begin{equation*}
=\frac{1}{n^{2}p^{2}}\sum_{i=1}^{n}\Gamma _{0}^{i}\left( \pi _{0}\cdot
I\otimes T+\left[ \sum_{\lambda =1}^{3}\pi _{\lambda }U_{\lambda }\right]
\otimes S\right) ^{n-k}\Gamma _{0}^{k-i}
\end{equation*}%
\begin{equation*}
=\frac{1}{n^{2}p^{2}}\sum_{i=1}^{n}\Gamma _{0}^{i}\left( \pi _{0}\cdot
I\otimes T+U\otimes S\right) ^{n-k}\Gamma _{0}^{k-i}
\end{equation*}%
\begin{equation*}
=\frac{1}{n^{2}p^{2}}\sum_{i=1}^{n}\Gamma _{0}^{i}\left( \sum_{p=0}^{n-k}\pi
_{0}^{p}\cdot U^{n-k-p}\otimes \left\{ T^{p},S^{n-k-p}\right\} \right)
\Gamma _{0}^{k-i}
\end{equation*}%
\begin{equation*}
=\frac{\left( p^{2}\right) ^{k/n}}{n^{2}p^{2}}\sum_{p=0}^{n-k}\pi
_{0}^{p}\cdot U^{n-k-p}\otimes \sum_{i=1}^{n}T^{i}\left\{
T^{p},S^{n-k-p}\right\} T^{k-i}.
\end{equation*}%
For $p<n-k\equiv l$ the last sum can be with the use of the relation (\ref%
{ga2}) modified%
\begin{equation}
\sum_{i=1}^{n}T^{i}\left\{ T^{p},S^{l-p}\right\} T^{k-i}=\left\{
T^{p},S^{l-p}\right\} T^{k}\sum_{i=1}^{n}\alpha ^{i\cdot (l-p)}  \label{s35}
\end{equation}%
\begin{equation*}
=\left\{ T^{p},S^{l-p}\right\} T^{k}\alpha ^{(l-p)}\frac{1-\alpha ^{n\cdot
(l-p)}}{1-\alpha ^{(l-p)}}=0,
\end{equation*}%
therefore only the term $p=n-k$ contributes: 
\begin{equation}
C_{k}=\frac{\left( p^{2}\right) ^{k/n}}{n^{2}p^{2}}\left( p_{0}^{2}\right)
^{(n-k)/n}n=\frac{1}{n}\left( \frac{p_{0}^{2}}{p^{2}}\right) ^{(n-k)/n}.
\label{s36}
\end{equation}%
So the sum (\ref{s32}) gives in total 
\begin{equation}
C=\frac{1}{n}\left[ 1+\left( \frac{p_{0}^{2}}{p^{2}}\right) ^{1/n}+\left( 
\frac{p_{0}^{2}}{p^{2}}\right) ^{2/n}+...+\left( \frac{p_{0}^{2}}{p^{2}}%
\right) ^{(n-1)/n}\right]  \label{s37}
\end{equation}%
\begin{equation*}
=\frac{1-p_{0}^{2}/p^{2}}{n\left[ 1-\left( p_{0}^{2}/p^{2}\right) ^{1/n}%
\right] },
\end{equation*}%
therefore Eq. (\ref{s27}) is satisfied with $Y,Y^{-1}$ given by Eq. (\ref%
{sa27}) and the proof is completed.

Solution of the Eq. (\ref{s26}) reads 
\begin{equation}
\Psi _{0}(p)=\left( 
\begin{array}{c}
\mathbf{0} \\ 
\cdot  \\ 
\mathbf{0} \\ 
\mathbf{g} \\ 
\mathbf{0} \\ 
\cdot  \\ 
\cdot  \\ 
\mathbf{0}%
\end{array}%
\right) ;\qquad \mathbf{g}\equiv \left( 
\begin{array}{c}
g_{1} \\ 
g_{2} \\ 
\cdot  \\ 
\cdot  \\ 
g_{n}%
\end{array}%
\right) ,\qquad \mathbf{0}\equiv \left( 
\begin{array}{c}
0 \\ 
0 \\ 
\cdot  \\ 
\cdot  \\ 
0%
\end{array}%
\right) ,  \label{s38}
\end{equation}%
i.e. the sequence of non zero components can be only in one block, whose
location depends on the choice of the phase of the power $(p^{2})^{1/n}$.
The $g_{j}$ are arbitrary functions of $p$ and simultaneously the constraint 
$p^{2}=m^{2}$ is required. Now we shall try to find the generators
satisfying the covariance condition for Eq. (\ref{s26}) 
\begin{equation}
\lbrack \mathbf{L_{\omega }},\Gamma _{0}(p)]=0  \label{s39}
\end{equation}%
together with the commutation relations (\ref{s19})-(\ref{s21}). Some hint
can be obtained from the Dirac equation transformed to the diagonal form in
an accordance with the relations (\ref{s27}), (\ref{sa27}). We shall use the
current representation of the Pauli and Dirac matrices 
\begin{equation}
\sigma _{1}=\left( 
\begin{array}{cc}
0 & 1 \\ 
1 & 0%
\end{array}%
\right) ,\qquad \sigma _{2}=\left( 
\begin{array}{cc}
0 & -i \\ 
i & 0%
\end{array}%
\right) ,\qquad \sigma _{3}=\left( 
\begin{array}{cc}
1 & 0 \\ 
0 & -1%
\end{array}%
\right) ,  \label{s40}
\end{equation}%
\begin{equation}
\gamma _{0}=\left( 
\begin{array}{cc}
\mathbf{1} & \mathbf{0} \\ 
\mathbf{0} & \mathbf{-1}%
\end{array}%
\right) ,\qquad \gamma _{j}=\left( 
\begin{array}{cc}
\mathbf{0} & \sigma _{j} \\ 
-\sigma _{j} & \mathbf{0}%
\end{array}%
\right) ;\qquad j=1,2,3,  \label{s41}
\end{equation}%
where the bold \textbf{0,1} stand for zero and unit matrices $2\times 2$.
The Dirac equation 
\begin{equation}
\left( \Gamma (p)-m\right) \Psi (p)=0,\qquad \Gamma (p)\equiv \sum_{\lambda
=0}^{3}p_{\lambda }\gamma _{\lambda }  \label{s42}
\end{equation}%
is covariant under the transformations generated by 
\begin{equation}
\mathbf{L_{\varphi _{j}}=}\frac{i}{4}\epsilon _{jkl}\gamma _{k}\gamma _{l}+i%
\frac{d}{d\varphi _{j}}=L_{\varphi _{j}}+i\frac{d}{d\varphi _{j}};\qquad
L_{\varphi _{j}}=\frac{1}{2}\left( 
\begin{array}{cc}
\sigma _{j} & \mathbf{0} \\ 
\mathbf{0} & \sigma _{j}%
\end{array}%
\right) ,  \label{s43}
\end{equation}%
\begin{equation}
\mathbf{L_{\psi _{j}}=}\frac{i}{2}\gamma _{0}\gamma _{j}+i\frac{d}{d\psi _{j}%
}=L_{\psi _{j}}+i\frac{d}{d\psi _{j}};\qquad L_{\psi _{j}}=\frac{i}{2}\left( 
\begin{array}{cc}
\mathbf{0} & \sigma _{j} \\ 
\sigma _{j} & \mathbf{0}%
\end{array}%
\right) ,  \label{s44}
\end{equation}%
where $j,k,l=1,2,3.$ Obviously, to preserve covariance, one has with the
transformation $\Gamma \rightarrow \Gamma _{0}=Y\Gamma Y^{-1}$ perform also 
\begin{equation}
\mathbf{L_{\omega }\rightarrow M_{\omega }}=Y(p)\mathbf{L_{\omega }}%
Y^{-1}(p).  \label{s45}
\end{equation}%
For the space rotations $\mathbf{L_{\varphi _{j}}\ }$commuting with both $%
\Gamma _{0},\Gamma $ and with the $Y$ from the relation (\ref{sa27}) the
result is quite straightforward 
\begin{equation}
\mathbf{M_{\varphi _{j}}=L_{\varphi _{j}}=}L_{\varphi _{j}}+i\frac{d}{%
d\varphi _{j}},  \label{s46}
\end{equation}%
i.e. the generators of the space rotations are not changed by the
transformation (\ref{s45}). The similar procedure with the Lorentz
transformations is slightly more complicated, nevertheless after calculation
of the commutator $[\mathbf{L_{\psi _{j}}},\Gamma _{0}/\sqrt{1+p_{0}/\sqrt{%
p^{2}}}]$ and a few further steps one obtains 
\begin{equation}
\mathbf{M_{\psi _{j}}=}M_{\psi _{j}}(p)+i\frac{d}{d\psi _{j}};\qquad M_{\psi
_{j}}(p)=\epsilon _{jkl}\frac{p_{k}L_{\varphi _{l}}}{p_{0}+\sqrt{p^{2}}}.
\label{s47}
\end{equation}%
So the generators (\ref{s46}), (\ref{s47}) guarantee the covariance of the
diagonalized Dirac equation obtained from Eq. (\ref{s42}) according to Lemma %
\ref{le4}. At the same time it is obvious, that having the set of generators 
$L_{\varphi _{j}}$ (with constant elements) of space rotations, one can
according to Eq. (\ref{s47}) construct the generators of Lorentz
transformations $M_{\psi _{j}}(p)$ (or $\mathbf{M_{\psi _{j}}}$), which
satisfy commutation relations (\ref{s19}) - (\ref{s21}). Obviously this
recipe is valid for any representation of infinitesimal space rotations. Let
us make a remark, that the algebra given by Eqs. (\ref{s46}), (\ref{s47})
appears in a slightly modified form already in \cite{foldy56}. Now, we shall
show, that if one requires a linear relation between the generators $M_{\psi
_{j}}$ and $L_{\varphi _{l}}$, like in Eq. (\ref{s47}), then this relation
can have a more general shape, than that in Eq. (\ref{s47}).

\begin{lemma}
\label{le5}Let $L_{\varphi _{j}}$ be matrices with constant elements
satisfying commutation relations (\ref{s19}). Then the operators 
\begin{equation}
\mathbf{M_{\psi _{j}}=}M_{\psi _{j}}(p)+i\frac{d}{d\psi _{j}};\qquad M_{\psi
_{j}}(p)=\frac{\kappa L_{\varphi _{j}}+\epsilon _{jkl}p_{k}L_{\varphi _{l}}}{%
p_{0}+\sqrt{p^{2}-\kappa ^{2}}},  \label{s48}
\end{equation}%
where $\kappa $ is any complex constant, satisfy the commutation relations (%
\ref{s20}), (\ref{s21}).
\end{lemma}

\noindent \textit{Proof:}

\noindent After insertion of the generators (\ref{s48}) into the relations (%
\ref{s20}), (\ref{s21}) one can check, that the commutation relations are
satisfied. In fact, it is sufficient to verify e.g. the commutators $[%
\mathbf{L_{\varphi _{1}}},\mathbf{L_{\psi _{2}}}],[\mathbf{L_{\varphi _{1}}},%
\mathbf{L_{\psi _{1}}}]$ and $[\mathbf{L_{\psi _{1}}},\mathbf{L_{\psi _{3}}}%
],$ the remaining follow from the cyclic symmetry.

Let us note, the formula (\ref{s48}) covers also the limit case $\left|
\kappa \right| \rightarrow \infty $, then 
\begin{equation}
M_{\psi _{j}}=iL_{\varphi _{j}}.  \label{s49}
\end{equation}%
On the other hand, the relation (\ref{s47}) corresponds to $\kappa =0.$ The
representations of the Lorentz group defined by the generators (\ref{s46}), (%
\ref{s48}) and differing only in the parameter $\kappa $ should be
equivalent in the sense 
\begin{equation}
\mathbf{M}_{\omega }(\kappa ^{\prime })=X^{-1}(p)\mathbf{M}_{\omega }(\kappa
)X(p).  \label{sa49}
\end{equation}%
We shall not make a general proof of this relation, but rather we shall
show, that the representations defined in the Lemma \ref{le5} and differing
only in $\kappa $, can be classified by the same mass $m^{2}=p^{2}$ and spin 
$\mathbf{s}^{2}=s(s+1).$ First, let us note, that the six generators
considered in the lemma together with the four generators $p_{\alpha }$ of
the space-time translations form the set of generators of the Poincar\'{e}
group. One can easily check, that the corresponding additional commutation
relations are satisfied: 
\begin{equation}
\lbrack p_{\alpha },p_{\beta }]=0,\qquad \lbrack \mathbf{M_{\varphi _{j}}}%
,p_{0}]=0,\qquad \lbrack p_{\alpha },\Gamma _{0}]=0,  \label{sa50}
\end{equation}%
\begin{equation}
\lbrack \mathbf{M_{\varphi _{j}}},p_{k}]=i\epsilon _{jkl}p_{l},\qquad
\lbrack \mathbf{M_{\psi _{j}}},p_{k}]=i\delta _{jk}p_{0},\qquad \lbrack 
\mathbf{M_{\psi _{j}}},p_{0}]=ip_{j}.  \label{sb50}
\end{equation}%
Further, the generators $\mathbf{M}_{\omega }$ can be rewritten in the
covariant notation 
\begin{equation}
\mathbf{M}_{jk}=\epsilon _{jkl}\mathbf{M}_{\varphi _{l}},\qquad \mathbf{M}%
_{j0}=\mathbf{M}_{\psi _{j}},\qquad \mathbf{M}_{\alpha \beta }=-\mathbf{M}%
_{\beta \alpha }.  \label{sb49}
\end{equation}%
Now the Pauli - Lubanski vector can be constructed: 
\begin{equation}
V_{\alpha }=\epsilon _{\alpha \beta \gamma \delta }\mathbf{M}^{\beta \gamma
}p^{\delta }/2,  \label{sc49}
\end{equation}%
which has satisfy 
\begin{equation}
V_{\alpha }V^{\alpha }=-m^{2}s(s+1),  \label{sd49}
\end{equation}%
where $s$ is the corresponding spin number. One can check, that after
inserting the generators (\ref{sb49}) into relations (\ref{sc49}), (\ref%
{sd49}), the result does not depend on $\kappa $%
\begin{equation*}
V_{\alpha }V^{\alpha }=-p^{2}\left( M_{\varphi _{1}}^{2}+M_{\varphi
_{2}}^{2}+M_{\varphi _{3}}^{2}\right) =-m^{2}s(s+1).
\end{equation*}

So the generators of the Lorentz group, which satisfy Eq. (\ref{s39}), can
have the form 
\begin{equation}
\mathbf{R_{\omega }}=R_{\omega }+i\frac{d}{d\omega };\ R_{\omega }=\left( 
\begin{array}{cccccc}
M_{\omega } & \qquad & \qquad & \qquad & \qquad &  \\ 
\qquad & M_{\omega } &  &  &  &  \\ 
&  & M_{\omega } &  &  &  \\ 
&  &  & \cdot &  &  \\ 
&  &  &  & \cdot &  \\ 
&  &  &  &  & M_{\omega }%
\end{array}%
\right) ,  \label{s50}
\end{equation}%
where $M_{\omega }$ are the $n\times n$ matrices defined in accordance with
the Lemma \ref{le5}. There are $n$ such matrices on the diagonal and
apparently these matrices may not be identical.

Finally, it is obvious that Eq. (\ref{s26}) is covariant also under any
infinitesimal transform 
\begin{equation}
\Lambda (d\xi )=I+id\xi \cdot K_{\xi },  \label{s51}
\end{equation}%
where the generators $K_{\xi }$ have the similar form as the generators (\ref%
{s50}) 
\begin{equation}
K_{\xi }=\left( 
\begin{array}{cccccc}
k_{\xi } & \qquad & \qquad & \qquad & \qquad &  \\ 
\qquad & k_{\xi } &  &  &  &  \\ 
&  & k_{\xi } &  &  &  \\ 
&  &  & \cdot &  &  \\ 
&  &  &  & \cdot &  \\ 
&  &  &  &  & k_{\xi }%
\end{array}%
\right)  \label{s52}
\end{equation}%
and generally their elements may depend on $p$. Obviously, one can put the
question: If the generators $L_{\varphi },k_{\xi }$ from Eqs. (\ref{s46}), (%
\ref{s52}) with constant elements represent the algebra of some group
(containing the rotation group as a subgroup), then what linear combination $%
M_{\psi _{j}}(p)$ of these generators satisfy the commutation relations (\ref%
{s20}), (\ref{s21}) for the generators of Lorentz transformations? In the
other words, what are the coefficients in summation 
\begin{equation}
M_{\psi _{j}}(p)=\sum_{k=1}^{3}c_{jk}(p)L_{\varphi _{k}}+\sum_{\xi }c_{j\xi
}k_{\xi }  \label{s53}
\end{equation}%
satisfying the commutation relations for the generators of the Lorentz
transformations? In this paper we shall not discuss this more general task,
for our present purpose it is sufficient, that we proved existence of the
generators of infinitesimal Lorentz transformations, under which the Eq. (%
\ref{s26}) is covariant.

\subsection{Finite transformations}

Now, having the infinitesimal transformations, one can proceed to finite
ones, corresponding to the parameters $\omega $ and $\xi $: 
\begin{equation}
\Psi _{0}^{\prime }(p^{\prime })=\Lambda (\omega )\Psi _{0}(p),\qquad \Psi
_{0}^{\prime }(p)=\Lambda (\xi )\Psi _{0}(p),  \label{s54}
\end{equation}%
where $p\rightarrow p^{\prime }$ is some of the transformations (\ref{sa10})
- (\ref{sd10}). The matrices $\Lambda $ satisfy 
\begin{equation}
\Lambda (\omega +d\omega )=\Lambda (\omega )\Lambda (d\omega ),\qquad
\Lambda (\xi +d\xi )=\Lambda (\xi )\Lambda (d\xi ),  \label{s55}
\end{equation}%
which for the parameters $\varphi $ (space rotations only) and $\xi $ imply 
\begin{equation}
\frac{d\Lambda (\varphi _{j})}{d\varphi _{j}}=i\Lambda (\varphi
_{j})R_{\varphi _{j}},\qquad \frac{d\Lambda (\xi )}{d\xi }=i\Lambda (\xi
)K_{\xi }.  \label{s56}
\end{equation}%
Assuming the constant elements of the matrices $R_{\varphi _{j}}$ and $%
K_{\xi }$, the solutions of the last equations can be written in the usual
exponential form 
\begin{equation}
\Lambda (\varphi _{j})=\exp (i\varphi _{j}R_{\varphi _{j}}),\qquad \Lambda
(\xi )=\exp (i\xi K_{\xi }).  \label{s57}
\end{equation}%
The space rotation by an angle $\varphi $ about the axis having the
direction $\vec{u},$ $\left| \vec{u}\right| =1$ is represented by 
\begin{equation}
\Lambda (\varphi ,\vec{u})=\exp \left[ i\varphi \left( \vec{u}\cdot \vec{R}%
_{\varphi }\right) \right] ;\qquad \vec{R}_{\varphi }=\left( R_{\varphi
_{1}},R_{\varphi _{2}},R_{\varphi _{3}}\right) .  \label{sa57}
\end{equation}%
For the Lorentz transformations we get instead of Eq. (\ref{s56}) 
\begin{equation}
\frac{d\Lambda (\psi _{j})}{d\psi _{j}}=if_{j}(\psi _{j})\Lambda (\psi
_{j})N_{j},  \label{s58}
\end{equation}%
where in accordance with Eqs. (\ref{sd10}) and (\ref{s48}) there stand for 
\begin{equation}
f_{j}(\psi _{j})=\frac{1}{p_{0}\cosh \psi _{j}+p_{j}\sinh \psi _{j}+\sqrt{%
p^{2}-\kappa ^{2}}},  \label{s59}
\end{equation}%
\begin{equation}
N_{j}=\kappa R_{\varphi _{j}}+\epsilon _{jkl}p_{k}R_{\varphi _{l}}.
\label{sa59}
\end{equation}%
The solution of Eq. (\ref{s58}) reads 
\begin{equation}
\Lambda (\psi _{j})=\exp \left( iF(\psi _{j})N_{j}\right) ;\qquad F(\psi
_{j})=\int_{0}^{\psi _{j}}f_{j}(\eta )d\eta .  \label{s60}
\end{equation}%
The Lorentz boost in a general direction $\vec{u}$ with the velocity $\beta $
is represented by 
\begin{equation}
\Lambda (\psi ,\vec{u})=\exp \left( iF(\psi )N\right) ,\qquad \tanh \psi
=\beta ,  \label{s61}
\end{equation}%
where 
\begin{equation}
F(\psi )=\int_{0}^{\psi }\frac{d\eta }{p_{0}\cosh \eta +\vec{p}\vec{u}\sinh
\eta +\sqrt{p^{2}-\kappa ^{2}}},  \label{s62}
\end{equation}%
\begin{equation}
N=\kappa \vec{u}\vec{R}_{\varphi }+\left( \vec{u}\times \vec{p}\right) \cdot 
\vec{R}_{\varphi }.  \label{sa62}
\end{equation}%
The corresponding integrals can be found e.g. in the handbook \cite{pru}.

Let us note, from the technical point of view, solution of the equation 
\begin{equation}
\frac{d\Lambda (t)}{dt}=\Omega (t)\Lambda (t),  \label{s63}
\end{equation}%
where $\Lambda ,\Omega $ are some square matrices, can be written in the
exponential form 
\begin{equation}
\Lambda (t)=\exp \left( \int_{0}^{t}\Omega (\eta )d\eta \right)  \label{s64}
\end{equation}%
only if the matrix $\Omega $ satisfies 
\begin{equation}
\left[ \Omega (t),\int_{0}^{t}\Omega (\eta )d\eta \right] =0.  \label{s65}
\end{equation}%
This condition is necessary for differentiation 
\begin{equation}
\frac{d\Lambda (t)}{dt}=\frac{d}{dt}\sum_{j=0}^{\infty }\frac{\left(
\int_{0}^{t}\Omega (\eta )d\eta \right) ^{j}}{j!}=\Omega
(t)\sum_{j=0}^{\infty }\frac{\left( \int_{0}^{t}\Omega (\eta )d\eta \right)
^{j}}{j!}  \label{s66}
\end{equation}%
\begin{equation*}
=\Omega (t)\Lambda (t)=\Lambda (t)\Omega (t).
\end{equation*}%
Obviously the condition (\ref{s66}) is satisfied for the generators of all
the considered transformations, including the Lorentz ones in Eq. (\ref{s61}%
), since the matrix $N$ does not depend on $\psi $. ($N$ depends only on the
momenta components perpendicular the direction of the Lorentz boost.)

\subsection{Equivalent transformations}

Now, from the symmetry of the Eq. (\ref{s26}) one can obtain the
corresponding transformations for the Eq. (\ref{s2}). The generators (\ref%
{s50}) satisfy the relations (\ref{s39}) and (\ref{s19}) - (\ref{s21}), it
follows that the generators%
\begin{equation}
\mathbf{R}_{\omega }(\Gamma )=Y^{-1}(p)\mathbf{R}_{\omega }(\Gamma
_{0})Y(p)=R_{\omega }(\Gamma )+i\frac{d}{d\omega },  \label{s67}
\end{equation}%
\begin{equation*}
R_{\omega }(\Gamma )=Y^{-1}(p)R_{\omega }(\Gamma _{0})Y(p)+iY^{-1}(p)\frac{%
dY(p)}{d\omega },
\end{equation*}%
where $\mathbf{R}_{\omega }(\Gamma _{0}),R_{\omega }(\Gamma _{0})$ are
generators (\ref{s50}) and $Y(p)$ is the transformation (\ref{sa27}), will
satisfy the same conditions, but with the relation (\ref{s17}) instead of
the relation (\ref{s39}). Similarly the generators $K_{\xi }(\Gamma _{0})$
in relation (\ref{s52}) will be for Eq. (\ref{s2}) replaced by 
\begin{equation}
K_{\xi }(\Gamma )=Y^{-1}(p)K_{\xi }(\Gamma _{0})Y(p).  \label{s68}
\end{equation}%
The finite transformations of the Eq. (\ref{s2}) and its solutions can be
obtained as follows. First let us consider the transformations $\Lambda
(\Gamma _{0},\omega ,\vec{u})$ given by Eqs. (\ref{sa57}) and (\ref{s61}).
In accordance with Eq. (\ref{s27}) we have 
\begin{equation}
\Gamma (p)=Y^{-1}(p)\Gamma _{0}(p)Y(p),\qquad \Gamma (p^{\prime
})=Y^{-1}(p^{\prime })\Gamma _{0}(p^{\prime })Y(p^{\prime })  \label{s69}
\end{equation}%
and correspondingly for the solutions of Eqs. (\ref{s2}), (\ref{s26}) 
\begin{equation}
\Psi (p)=Y^{-1}(p)\Psi _{0}(p),\qquad \Psi ^{\prime }(p^{\prime
})=Y^{-1}(p^{\prime })\Psi _{0}^{\prime }(p^{\prime }).  \label{s70}
\end{equation}%
Since 
\begin{equation}
\Psi _{0}^{\prime }(p^{\prime })=\Lambda (\Gamma _{0},\omega ,\vec{u})\Psi
_{0}(p),  \label{s71}
\end{equation}%
then Eqs. (\ref{s70}) imply 
\begin{equation}
\Psi ^{\prime }(p^{\prime })=\Lambda (\Gamma ,\omega ,\vec{u})\Psi
(p);\qquad \Lambda (\Gamma ,\omega ,\vec{u})=Y^{-1}(p^{\prime })\Lambda
(\Gamma _{0},\omega ,\vec{u})Y(p).  \label{s72}
\end{equation}%
Similarly, the transformations $\Lambda (\Gamma _{0},\xi )$ given by Eq. (%
\ref{s57}) are for Eq. (\ref{s2}) replaced by 
\begin{equation}
\Lambda (\Gamma ,\xi )=Y^{-1}(p)\Lambda (\Gamma _{0},\xi )Y(p).  \label{s73}
\end{equation}%
Let us note, all the symmetries of Eq. (\ref{s2}) like the transformation (%
\ref{s73}), which are not connected with a change of the reference frame ($p$%
), can be in accordance with the relation (\ref{s25}) expressed 
\begin{equation}
\Lambda (\Gamma ,X)=Z(\Gamma ,X,\Gamma ),  \label{s74}
\end{equation}%
where $Z$ is defined by Eq. (\ref{s22}) and $X(p)$ is any matrix for which
there exists $Z(\Gamma ,X,\Gamma )^{-1}$. Further, it is obvious that if we
have some set of generators $\mathbf{R}_{\omega }(\Gamma )$ satisfying the
relations (\ref{s17}) and (\ref{s19}) - (\ref{s21}), then also any set 
\begin{equation}
\mathbf{\hat{R}}_{\omega }(\Gamma )=Z(\Gamma ,X,\Gamma )^{-1}\mathbf{R}%
_{\omega }(\Gamma )Z(\Gamma ,X,\Gamma )  \label{s75}
\end{equation}%
satisfies these conditions. For the finite transformations one gets
correspondingly 
\begin{equation}
\hat{\Lambda}(\Gamma ,\omega ,\vec{u})=Z\left( \Gamma (p^{\prime
}),X(p^{\prime }),\Gamma (p^{\prime })\right) ^{-1}\Lambda (\Gamma ,\omega ,%
\vec{u})Z\left( \Gamma (p),X(p),\Gamma (p)\right) .  \label{s76}
\end{equation}%
In the same way, the sets of equivalent generators and transformations can
be obtained for the diagonalized equation (\ref{s26}).

Let us remark, according to the Lemma \ref{le4} there exists the set of
transformations $\Gamma (p)\leftrightarrow \Gamma _{0}(p)$ given by the
relation (\ref{s27}). We used its particular form (\ref{sa27}), but how will
the generators 
\begin{equation}
\mathbf{R}_{\omega }(\Gamma ,X_{k})=Z(\Gamma _{0},X_{k},\Gamma )^{-1}\mathbf{%
R}_{\omega }(\Gamma _{0})Z(\Gamma _{0},X_{k},\Gamma );\qquad k=1,2
\label{s77}
\end{equation}%
differ for the two different matrices $X_{1}$ and $X_{2}$? The last relation
implies 
\begin{equation}
\mathbf{R}_{\omega }(\Gamma ,X_{1})=Z(\Gamma _{0},X_{1},\Gamma
)^{-1}Z(\Gamma _{0},X_{2},\Gamma )\mathbf{R}_{\omega }(\Gamma
,X_{2})Z(\Gamma _{0},X_{2},\Gamma )^{-1}Z(\Gamma _{0},X_{1},\Gamma )
\label{s78}
\end{equation}%
and according to the relation (\ref{s23})%
\begin{equation}
\Gamma Z(\Gamma _{0},X_{1},\Gamma )^{-1}Z(\Gamma _{0},X_{2},\Gamma
)=Z(\Gamma _{0},X_{1},\Gamma )^{-1}\Gamma _{0}Z(\Gamma _{0},X_{2},\Gamma )
\label{s79}
\end{equation}%
\begin{equation*}
=Z(\Gamma _{0},X_{1},\Gamma )^{-1}Z(\Gamma _{0},X_{2},\Gamma )\Gamma ,
\end{equation*}%
which means 
\begin{equation}
\lbrack Z(\Gamma _{0},X_{1},\Gamma )^{-1}Z(\Gamma _{0},X_{2},\Gamma ),\Gamma
]=0.  \label{s80}
\end{equation}%
It follows that there must exists a matrix $X_{3}$ [e.g. according to
implication (\ref{sa25}) one can put $X_{3}=Z(\Gamma _{0},X_{1},\Gamma
)^{-1}Z(\Gamma _{0},X_{2},\Gamma )$] so that 
\begin{equation}
Z(\Gamma ,X_{3},\Gamma )=Z(\Gamma _{0},X_{1},\Gamma )^{-1}Z(\Gamma
_{0},X_{2},\Gamma ),  \label{s81}
\end{equation}%
then the relation (\ref{s78}) can be rewritten 
\begin{equation}
\mathbf{R}_{\omega }(\Gamma ,X_{1})=Z(\Gamma ,X_{3},\Gamma )\mathbf{R}%
_{\omega }(\Gamma ,X_{2})Z(\Gamma ,X_{3},\Gamma )^{-1},  \label{s82}
\end{equation}%
i.e. the generators $\mathbf{R}_{\omega }(\Gamma ,X_{1}),\mathbf{R}_{\omega
}(\Gamma ,X_{2})$ are equivalent in the sense of the relation (\ref{s75}).

\subsection{Scalar product and unitary representations}

\begin{definition}
The scalar product of the two functions satisfying Eq. (\ref{s2}) or (\ref%
{s26}) is defined: 
\begin{equation}
\left( \Phi (p),\Psi (q)\right) =\left\{ 
\begin{array}{c}
0 \\ 
\Phi ^{\dagger }(p)W(p)\Psi (q)%
\end{array}%
\right. \quad \mathrm{for}\quad 
\begin{array}{c}
p\neq q \\ 
p=q%
\end{array}%
,  \label{s94}
\end{equation}%
where the metric $W$ is the matrix, which satisfies%
\begin{equation}
W^{\dagger }(p)=W(p),  \label{s95}
\end{equation}%
\begin{equation}
R_{\omega }^{\dagger }(p)W(p)-W(p)R_{\omega }(p)+i\frac{dW}{d\omega }=0,
\label{s96}
\end{equation}%
\begin{equation}
K_{\xi }^{\dagger }(p)W(p)-W(p)K_{\xi }(p)=0.  \label{s97}
\end{equation}
\end{definition}

\noindent The conditions (\ref{s96}), (\ref{s97}) in the above definition
imply, that the scalar product is invariant under corresponding
infinitesimal transformations. For example for the Lorentz group the
transformed scalar product reads 
\begin{equation}
\Phi ^{\prime \dagger }(p^{\prime })W(p^{\prime })\Psi ^{\prime }(p^{\prime
})  \label{s98}
\end{equation}%
\begin{equation*}
=\Phi ^{\dagger }(p)\left( I-id\omega R_{\omega }^{\dagger }(p)\right)
\left( W(p)+d\omega \frac{dW}{d\omega }\right) \left( I+id\omega R_{\omega
}(p)\right) \Psi (p)
\end{equation*}%
and with the use of the condition (\ref{s96}) one gets 
\begin{equation}
\Phi ^{\prime \dagger }(p^{\prime })W(p^{\prime })\Psi ^{\prime }(p^{\prime
})=\Phi ^{\dagger }(p)W(p)\Psi (p).  \label{s99}
\end{equation}%
According to a general definition, the transformations conserving the scalar
product are unitary. In this way the Eqs. (\ref{s96}), (\ref{s97}) represent
the condition of unitarity for representation of the corresponding group
generated by $\mathbf{R}_{\omega }$ and $K_{\xi }.$

How to choose these generators and the matrix $W(p)$ to solve Eqs. (\ref{s96}%
), (\ref{s97})? Similarly as in the case of solution of Eqs. (\ref{s17}) and
(\ref{s19}) - (\ref{s21}) it is convenient to begin with the representation
related to the canonical equation (\ref{s26}). Apparently the generators of
the space rotations can be chosen Hermitian 
\begin{equation}
R_{\varphi _{j}}^{\dagger }(\Gamma _{0})=R_{\varphi _{j}}(\Gamma _{0}).
\label{s102}
\end{equation}%
Then also for the Lorentz transformations one gets 
\begin{equation}
R_{\psi _{j}}^{\dagger }(\Gamma _{0})=R_{\psi _{j}}(\Gamma _{0}),
\label{s103}
\end{equation}%
provided that the constant $\kappa $ in Eq. (\ref{s48}) is real and $\left|
\kappa \right| \leq m$. Also the generators $K_{\xi }$ can be chosen in the
same way: 
\begin{equation}
K_{\xi }^{\dagger }(\Gamma _{0})=K_{\xi }(\Gamma _{0}).  \label{s104}
\end{equation}%
It follows, that instead of the conditions (\ref{s96}), (\ref{s97}) one can
write 
\begin{equation}
\lbrack \mathbf{R}_{\omega }(\Gamma _{0}),W(\Gamma _{0})]=0,\qquad \lbrack
K_{\xi }(\Gamma _{0}),W(\Gamma _{0})]=0.  \label{s105}
\end{equation}%
The structure of the generators $\mathbf{R}_{\omega }(\Gamma _{0}),K_{\xi
}(\Gamma _{0})$ given by Eqs. (\ref{s50}), (\ref{s52}) suggests, that the
metric $W$ satisfying the condition (\ref{s105}) can have a similar
structure, but in which the corresponding blocks on the diagonal are
occupied by unit matrices multiplied by some constants. Nevertheless, let us
note, that the condition (\ref{s105}) in general can be satisfied also for
some other structures of $W(\Gamma _{0})$.

From $W(\Gamma _{0})$ we can obtain matrix $W(\Gamma )-$ the metric for the
scalar product of the two solutions of Eq. (\ref{s2}). One can check, that
after the transformations 
\begin{equation}
\mathbf{R}_{\omega }(\Gamma _{0})\rightarrow \mathbf{R}_{\omega }(\Gamma
,X)=Z(\Gamma _{0},X,\Gamma )^{-1}\mathbf{R}_{\omega }(\Gamma _{0})Z(\Gamma
_{0},X,\Gamma ),  \label{s106}
\end{equation}%
\begin{equation}
K_{\xi }(\Gamma _{0})\rightarrow K_{\xi }(\Gamma ,X)=Z(\Gamma _{0},X,\Gamma
)^{-1}K_{\xi }(\Gamma _{0})Z(\Gamma _{0},X,\Gamma )  \label{s107}
\end{equation}%
and simultaneously 
\begin{equation}
W(\Gamma _{0})\rightarrow W(\Gamma ,X)=Z(\Gamma _{0},X,\Gamma )^{\dagger
}W(\Gamma _{0})Z(\Gamma _{0},X,\Gamma )  \label{s108}
\end{equation}%
the unitarity in the sense of conditions (\ref{s96}), (\ref{s97}) is
conserved - in spite of the fact that equalities (\ref{s102}) - (\ref{s104})
may not hold for $R_{\omega }(\Gamma ,X),K_{\xi }(\Gamma ,X)$.

\subsection{Space-time representation and Green functions}

If we take the solutions of the wave equation (\ref{s2}) or (\ref{s26}) in
the form of the functions $\Psi (p)$, for which there exists the Fourier
picture 
\begin{equation}
\tilde{\Psi}(x)=\frac{1}{(2\pi )^{4}}\int \Psi (p)\delta (p^{2}-m^{2})\exp
(-ipx)d^{4}p,  \label{s109}
\end{equation}%
then the space of the functions $\tilde{\Psi}(x)$ constitutes the $x-$
representation of wave functions. Correspondingly, for all the operators $%
D(p)$ given in the $p-$ representation and discussed in the previous
paragraphs, one can formally define their $x-$ representation:

\begin{equation}
\tilde{D}(z)=\frac{1}{(2\pi )^{4}}\int D(p)\exp \left( -ipz\right) d^{4}p,
\label{s111}
\end{equation}%
which means%
\begin{equation}
\tilde{D}\tilde{\Psi}(x)=\frac{1}{(2\pi )^{4}}\int D(p)\Psi (p)\delta
(p^{2}-m^{2})\exp (-ipx)d^{4}p  \label{s112}
\end{equation}%
\begin{equation*}
=\frac{1}{(2\pi )^{4}}\int D(p)\exp (-ipx)\tilde{\Psi}(y)\exp
(ipy)d^{4}yd^{4}p=\frac{1}{(2\pi )^{4}}\int \tilde{D}(x-y)\tilde{\Psi}%
(y)d^{4}y.
\end{equation*}%
In this way we get \ for our operators: 
\begin{equation}
\Gamma _{0}(p)\rightarrow \tilde{\Gamma}_{0}(z)=Q_{0}\frac{1}{(2\pi )^{4}}%
\int (p^{2})^{1/n}\exp (-ipz)d^{4}p;\quad pz\equiv p_{0}z_{0}-\vec{p}\vec{z},
\label{s83}
\end{equation}%
\begin{equation}
\Gamma (p)\rightarrow \tilde{\Gamma}(z)=\sum_{\lambda =0}^{3}Q_{\lambda }%
\frac{1}{(2\pi )^{4}}\int p_{\lambda }^{2/n}\exp (-ipz)d^{4}p,  \label{s84}
\end{equation}%
\begin{equation}
\mathbf{R}_{\varphi _{j}}(\Gamma _{0})\rightarrow \mathbf{\tilde{R}}%
_{\varphi _{j}}(\Gamma _{0})=R_{\varphi _{j}}(\Gamma _{0})+i\frac{d}{d\tilde{%
\varphi}_{j}};\qquad \frac{d}{d\tilde{\varphi}_{j}}=-\epsilon _{jkl}x_{k}%
\frac{\partial }{\partial x_{l}},  \label{s85}
\end{equation}%
\begin{equation}
\mathbf{R}_{\psi _{j}}(\Gamma _{0})\rightarrow \mathbf{\tilde{R}}_{\psi
_{j}}(z)  \label{s86}
\end{equation}%
\begin{equation*}
=\frac{1}{(2\pi )^{4}}\int \frac{\kappa R_{\varphi _{j}}(\Gamma
_{0})+\epsilon _{jkl}p_{k}R_{\varphi _{l}}(\Gamma _{0})}{p_{0}+\sqrt{%
p^{2}-\kappa ^{2}}}\exp (-ipz)d^{4}p+i\frac{d}{d\tilde{\psi}_{j}};
\end{equation*}%
\begin{equation*}
\frac{d}{d\tilde{\psi}_{j}}=-x_{0}\frac{\partial }{\partial x_{j}}-x_{j}%
\frac{\partial }{\partial x_{0}}
\end{equation*}%
and in the same way 
\begin{equation}
\mathbf{R}_{\omega }(\Gamma )\rightarrow \mathbf{\tilde{R}}_{\omega }(z)
\label{s87}
\end{equation}%
\begin{equation*}
=\frac{1}{(2\pi )^{4}}\int Z(\Gamma _{0},X,\Gamma )^{-1}\mathbf{R}_{\omega
}(\Gamma _{0})Z(\Gamma _{0},X,\Gamma )\exp (-ipz)d^{4}p.
\end{equation*}%
Apparently, the similar relations are valid also for the remaining operators 
$K_{\xi },W,Z,Z^{-1}$ and the finite transformations $\Lambda $ in the $x-$%
representation. Concerning the translations, the usual correspondence is
valid: $p_{\alpha }\rightarrow i\partial _{\alpha }$.

Further, the solutions of the inhomogeneous version of the Eqs. (\ref{s26}),
(\ref{s2}) 
\begin{equation}
\left( \Gamma _{0}(p)-\mu \right) G_{0}(p)=I,\qquad \left( \Gamma (p)-\mu
\right) G(p)=I  \label{s88}
\end{equation}%
can be obtained with the use of formula (\ref{gaa23}): 
\begin{equation}
G_{0}(p)=\frac{(\Gamma _{0}-\alpha \mu )(\Gamma _{0}-\alpha ^{2}\mu
)...(\Gamma _{0}-\alpha ^{n-1}\mu )}{p^{2}-m^{2}},  \label{s89}
\end{equation}%
\begin{equation}
G(p)=\frac{(\Gamma -\alpha \mu )(\Gamma -\alpha ^{2}\mu )...(\Gamma -\alpha
^{n-1}\mu )}{p^{2}-m^{2}}  \label{sa89}
\end{equation}%
and Eq. (\ref{s27}) implies 
\begin{equation}
G(p)=Z(\Gamma _{0},X,\Gamma )^{-1}G_{0}(p)Z(\Gamma _{0},X,\Gamma ).
\label{s90}
\end{equation}%
Apparently, the functions 
\begin{equation}
\tilde{G}_{0}(x)=\frac{1}{(2\pi )^{4}}\int G_{0}(p)\exp (-ipx)d^{4}p,
\label{sa90}
\end{equation}%
\begin{equation}
\tilde{G}(x)=\frac{1}{(2\pi )^{4}}\int G(p)\exp (-ipx)d^{4}p  \label{s91}
\end{equation}%
formally satisfy Eqs. (\ref{s88}) in the $x-$representation 
\begin{equation}
\int \tilde{\Gamma}_{0}(x-y)\tilde{G}_{0}(y)d^{4}y-\mu \tilde{G}%
_{0}(x)=\delta ^{4}(x),  \label{s92}
\end{equation}%
\begin{equation}
\int \tilde{\Gamma}(x-y)\tilde{G}(y)d^{4}y-\mu \tilde{G}(x)=\left[
\sum_{\lambda =0}^{3}Q_{\lambda }(i\partial _{\lambda })^{2/n}-\mu \right] 
\tilde{G}(x)=\delta ^{4}(x).  \label{s93}
\end{equation}%
The last equation contains the fractional derivatives defined in \cite{zav}.
Obviously, the functions $\tilde{G}_{0},\tilde{G}$ can be identified with
the Green functions related to $x-$ representation of Eqs. (\ref{s26}), (\ref%
{s2}).

With the exception of the operators $\mathbf{\tilde{R}}_{\varphi
_{j}}(\Gamma _{0}),\tilde{W}(\Gamma _{0})$ and $i\partial _{\alpha }$ all
the remaining operators considered above are pseudo-differential ones, which
are in general non-local. The ways, how to deal with such operators, are
suggested in \cite{lam},\cite{bar},\cite{zav}. A more general treatise of
the pseudo-differential operators can be found e.g. in \cite{jos}-\cite{tre}%
. In our case it is significant, that the corresponding integrals will
depend on the choice of passing about the singularities and the choice of
the cuts of the power functions $p^{2j/n}$. This choice should reflect
contained physics, however corresponding discussion would exceed the scope
of this paper.

\section{Summary and concluding remarks}

In this paper we have first studied the algebra of the matrices $%
Q_{pr}=S^{p}T^{r}$ generated by the pair of matrices $S,T$ with the
structure given by the \textit{Definition \ref{def1}}. We have proved, that
for a given $n\geq 2$ one can in the corresponding set $\{Q_{pr}\}$ always
find a triad, for which Eq. (\ref{ga37}) is satisfied, hereat the Pauli
matrices represent its particular case $n=2$. On this base we have got the
rule, how to construct the generalized Dirac matrices [Eqs. (\ref{ga39}), (%
\ref{gb39})]. Further we have shown, that there is a simple relation [Eqs. (%
\ref{ga40}), (\ref{ga41})] between the set $\{Q_{pr}\}$ and the algebra of
generators of the fundamental representation of the $SU(n)$ group.

In the further part, using the generalized Dirac matrices we have
demonstrated, how one can from the roots of the D'Alambertian operator
generate a class of relativistic equations containing the Dirac equation as
a particular case. In this context we have shown, how the corresponding
representations of the Lorentz group, which guarantee the covariance of
these equations, can be found. At the same time we have found additional
symmetry transformations on these equations. Further, we have suggested how
one can define the scalar product in the space of the corresponding wave
functions and make the unitary representation of the whole group of
symmetry. Finally, we have suggested, how to construct the corresponding
Green functions. In the $x-$ representation the equations themselves and all
the mentioned transformations are in general non-local, being represented by
the fractional derivatives and pseudo-differential operators in the four
space-time dimensions.

In line with the choice of the representation of the rotation group used for
the construction of the unitary representation of the Lorentz group
according to which the equations transform, one can ascribe to the related
wave functions the corresponding spin - and further quantum numbers
connected with the additional symmetries. Nevertheless it is obvious, that
before more serious physical speculations, one should answer some more
questions requiring a further study. Perhaps the first could be the problem
how to introduce the interaction. The usual direct replacement%
\begin{equation*}
\partial _{\lambda }\rightarrow \partial _{\lambda }+igA_{\lambda }(x)
\end{equation*}%
would lead to the difficulties, first of all with the rigorous definition of
the terms like%
\begin{equation*}
\left( \partial _{\lambda }+igA_{\lambda }(x)\right) ^{2/n}.
\end{equation*}%
On the end one should answer the more general question: Is it possible on
the base of the discussed wave equations to build up a meaningful quantum
field theory? \ \\[5mm]
\noindent \textbf{Acknowledgement:} I would like to thank J.Tolar and M.
Rausch de Traubenberg for drawing my attention to the cited articles on
generalized Clifford algebras and M. Bedn\'{a}\v{r} for critical reading of
the manuscript.

\end{document}